\documentclass[sigconf]{acmart}

\AtBeginDocument{%
  \providecommand\BibTeX{{%
    \normalfont B\kern-0.5em{\scshape i\kern-0.25em b}\kern-0.8em\TeX}}}

\setcopyright{acmcopyright}
\copyrightyear{2023}
\acmYear{2023}

\copyrightyear{2024}
\acmYear{2024}
\setcopyright{rightsretained}
\acmConference[ICSE '24]{2024 IEEE/ACM 46th International Conference on Software Engineering}{April 14--20, 2024}{Lisbon, Portugal}
\acmBooktitle{2024 IEEE/ACM 46th International Conference on Software Engineering (ICSE '24), April 14--20, 2024, Lisbon, Portugal}
\acmDOI{10.1145/3597503.3639175}
\acmISBN{979-8-4007-0217-4/24/04}

\usepackage{amsmath}

\usepackage{amssymb,amsfonts}
\usepackage{graphicx}
\usepackage{textcomp}
\usepackage{xcolor}

\usepackage{pifont}
\usepackage{multirow}
\usepackage{subfig}
\usepackage{graphicx}
\usepackage[ruled,linesnumbered]{algorithm2e}
\usepackage{algorithmic}
\usepackage{hyperref}
\usepackage{amsfonts}
\usepackage{comment}
\usepackage{todonotes}

\newcommand{\cat}{$\mathtt{CAT}$\xspace} 
\newcommand{\framework}{$\mathtt{ACAV}$\xspace} 

\begin{document}

\title{ACAV: A Framework for Automatic Causality Analysis in Autonomous Vehicle Accident Recordings}
\author{Huijia Sun}
\orcid{0009-0000-8504-2733}
\affiliation{
\institution{ShanghaiTech University}
\country{China}
}
\email{sunhj2022@shanghaitech.edu.cn}

\author{Christopher M. Poskitt}
\orcid{0000-0002-9376-2471}
\affiliation{
\institution{Singapore Management University}
\country{Singapore}}
\email{cposkitt@smu.edu.sg}

\author{Yang Sun}
\orcid{0000-0002-2409-2160}
\affiliation{
\institution{Singapore Management University}
\country{Singapore}
}
\email{yangsun.2020@phdcs.smu.edu.sg}

\author{Jun Sun}
\orcid{0000-0002-3545-1392}
\affiliation{
\institution{Singapore Management University}
\country{Singapore}}
\email{junsun@smu.edu.sg}

\author{Yuqi Chen}
\orcid{0000-0003-2988-6012}
\authornote{Yuqi Chen is the corresponding author.}
\affiliation{
\institution{ShanghaiTech University}
\country{China}
}
\email{chenyq@shanghaitech.edu.cn}

\begin{abstract}
The rapid progress of autonomous vehicles~(AVs) has brought the prospect of a driverless future closer than ever.
Recent fatalities, however, have emphasized the importance of safety validation through large-scale testing.
Multiple approaches achieve this fully automatically using high-fidelity simulators, i.e., by generating diverse driving scenarios and evaluating autonomous driving systems~(ADSs) against different test oracles.
While effective at finding violations, these approaches do not identify the decisions and actions that \emph{caused} them---information that is critical for improving the safety of ADSs.
To address this challenge, we propose \framework, an automated framework designed to conduct causality analyses for AV accident recordings in two stages.
First, we apply feature extraction schemas based on the messages exchanged between ADS modules, and use a weighted voting method to discard frames of the recording unrelated to the accident.
Second, we use safety specifications to identify safety-critical frames and deduce causal events by applying \cat---our causal analysis tool---to a station-time graph.
We evaluated \framework on the Apollo ADS, finding that it can identify five distinct types of causal events in 93.64\% of 110 accident recordings generated by an AV testing engine. We further evaluated \framework on 1206 accident recordings collected from versions of Apollo injected with specific faults, finding that it can correctly identify causal events in 96.44\% of the accidents triggered by prediction errors, and 85.73\% of the accidents triggered by planning errors. 
\end{abstract}



\maketitle

\section{Introduction}

Autonomous Vehicles (AVs) are set to bring about a paradigm shift in transportation.
AVs operate through the use of advanced Autonomous Driving Systems (ADSs), which eliminate the need for human drivers to control the vehicle's movements.
ADSs are considered highly security-critical systems, as malfunctions can result in severe consequences \cite{bay_city_news, fatal_tesla_crash, dixit2016autonomous}.
For example, a minor error in trajectory prediction can lead to potentially hazardous or even fatal situations for passengers, other road users, and pedestrians.
Thus, it is imperative for AV developers to subject ADSs to rigorous testing to ensure their accuracy and reliability.
Given that on-road testing suffers from several limitations (such as safety risks and high expenses), simulation-based testing in high-fidelity simulators such as SVL~\cite{rong2020lgsvl} and CARLA~\cite{dosovitskiy2017carla} has emerged as a popular approach for evaluating AVs.

Many researchers utilize search-based~\cite{althoff2017commonroad, dreossi2019compositional, abdessalem2018testing_mos, riccio2020model, zohdinasab2021deephyperion, abdessalem2018testing} and sampling techniques~\cite{birkemeyer2022feature, feng2021intelligent, zhao2016accelerated, huang2017accelerated, zhao2017accelerated, wang2021combining} to generate and execute test cases against a set of testing oracles in simulation environments.
This provides a controlled and repeatable means of evaluating AVs without the risks associated with real-world testing.
For instance, AV-Fuzzer~\cite{li2020av} uses fuzzing to generate scenarios that cause safety violations such as near- and actual collisions.
LawBreaker~\cite{sun2022lawbreaker}, also based on fuzzing, further evaluates AVs against specifications of national traffic laws (e.g., rules for crossing junctions).
While these methods are effective at finding different violations, they typically do not provide insight into the specific decisions and actions of the AV that ultimately \emph{caused} the violations.
Such information is critical for engineers to improve the safety and reliability of AVs but is time-consuming and laborious to extract manually, especially in large-scale testing frameworks.
This problem has been emphasized in a recent study~\cite{10.1145/3540250.3549111}: given the vast amounts of driving recordings collected during testing, there is an urgent need for automated tools to support ADS engineers, e.g., ~in tasks such as clipping and interpreting. 

Causality analysis has been proposed within the software engineering community as a means to assist developers in deducing the underlying causes of faulty behaviors observed in a failed test case.
This technique has shown notable effectiveness in analyzing complex systems~\cite{zhang2017transfer, correa2017causal, correa2018generalized, bareinboim2015recovering}.
Unfortunately, given an AV accident recording extracted from a simulator, it is non-trivial to apply existing causality analysis techniques due to two main challenges.
First, ADSs consist of multiple independent, decoupled modules that communicate via message passing.
Thus, minor faults in one module can eventually propagate into serious faults in other modules.
For instance, an incorrect trajectory prediction may be used by an AV's planning module in a way that leads to an accident: in this context, the planning module is not solely to blame.
Second, the analysis space of a typical accident recording is huge, requiring new approaches for identifying the accident-related segments that should be focused on.

\begin{figure*}
\centering
\centerline{\includegraphics[width=1.0\linewidth]{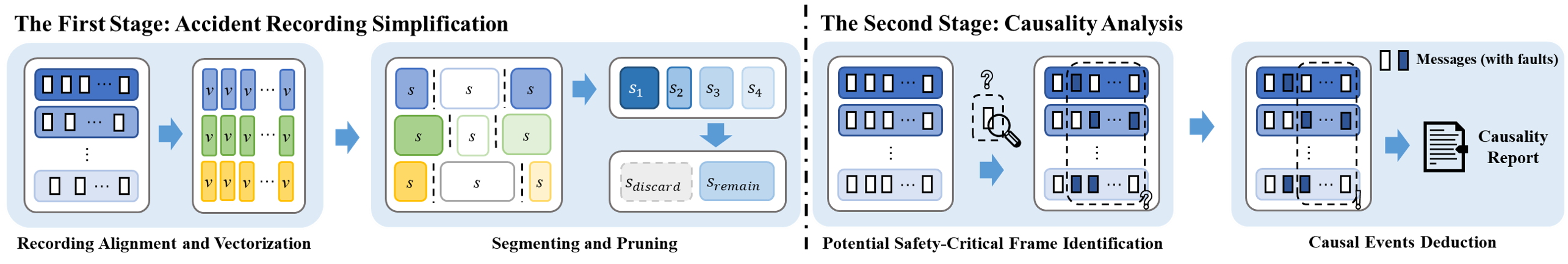}}
\caption{Overview of \framework: the first stage vectorizes data exchanged between ADS modules and discards recording segments irrelevant to the accident; the second stage performs a causality analysis using the \cat tool}
\label{fig:overview}
\end{figure*}

To address these challenges, we present \framework, a framework for Automatic Causality analysis of AV accident recordings.
Our approach consists of two stages: \emph{accident recording simplification} and \emph{causality analysis}, summarized in the high-level workflow diagram of Figure~\ref{fig:overview}.
In the first stage, we define and apply feature extraction schemas based on the messages exchanged between ADS modules.
These schemas are used to vectorize information about the map, as well as the AV's perception, prediction, and planning.
We then propose a weighted voting method to integrate the slicing plans generated by these schemas, allowing for segments unrelated to the safety violation to be discarded.
In the second stage, we identify safety-critical frames using an a priori method based on safety specifications extracted from the driver's handbook and traffic laws of California.
Next, we apply our novel causality analysis tool, \cat, to identify the causal events of an accident by analyzing the Station-Time graph (ST graph).
In conclusion, our framework is designed to identify the safety-critical frames that brought about an accident, and to generate detailed reports enumerating potential causes. This functionality empowers engineers to gain a comprehensive understanding of the accident dynamics without first needing to replay entire recordings. 

To evaluate the effectiveness of our framework, we implemented it for Apollo 7.0 \cite{apollo70} and the SVL simulator~\cite{rong2020lgsvl}, which are widely used tools in the field of autonomous driving research and development.
Using an AV testing engine~\cite{sun2022lawbreaker}, we collected a total of 110 accident recordings, including accidents involving intersections, merging, and tailgating.
We applied \framework to vectorize and simplify these recordings, finding that \framework achieved a 62.23\% reduction ratio rate without discarding critical frames, demonstrating its effectiveness in simplifying accident recordings.
Upon analyzing the simplified recordings with \cat, our approach identifies five distinct types of causal events in 93.64\% of the recordings, including incorrect priority prediction (found 26 times), incorrect trajectory prediction (51 times), improper behavioral planning (17 times), unsafe motion planning (67 times), and vehicle out-of-control (103 times).
Finally, we further evaluated \framework on 1206 accident recordings collected from versions of Apollo injected with specific faults, finding that it can correctly identify causal events in 96.44\% of the accidents triggered by prediction errors, and 85.73\% of the accidents triggered by planning errors.

Our website~\cite{ourweb} provides videos of multiple accidents involving the Apollo ADS, together with the complete accident reports generated by \framework, as well as our source code.

Overall, we make the following contributions:

\begin{itemize}
    \item Feature extraction schemas for vectorizing map, perception, prediction, and planning information from ADS messages in AV accident recordings.
    \item A mechanism for identifying and discarding recording segments unrelated to the accident.
    \item A tool for identifying safety-critical frames from an accident recording by leveraging ST graphs.
    \item \framework, which to the best of our knowledge, is the first modular framework for AV accident analysis and explanation.
    \item An implementation for Apollo 7.0 and SVL that is able to identify five types of causal faults in AV accident recordings.
\end{itemize}

The paper is organized as follows.
In Section~\ref{sec:background}, we review some essential background and present a motivational example. Section~\ref{sec:framework_design} introduces the design of \framework, including the detailed algorithms of its two stages.
Section~\ref{sec:evaluation} evaluates whether \framework achieves its goal of identifying causal events from AV accident recordings.
Finally, Section~\ref{sec:related_work} compares our approach against some related work, before Section~\ref{sec:conclusion} concludes.

\section{Background and Example}
\label{sec:background}

\subsection{Multi-Module ADSs}

The ADSs of AVs are composed of various modules, including perception, localization, prediction, planning, and control. 
These modules utilize multiple sensors, such as cameras, LiDAR, GNSS, and IMU, that capture raw data (e.g., images, 3D point clouds) about the AV's state as well as the environment it is operating in.
To facilitate collaboration among the modules, industrial-level ADSs use a publish-subscribe (i.e., message-based) model for communication. 
Each module subscribes to one or more channels in the ADS to obtain the required inputs and publishes its output as a message to the corresponding channels.

Specifically, the localization module constantly processes data collected from the GPS, IMU, and (sometimes) LiDAR, then publishes messages containing information about the vehicle's position, orientation, and speed.
The perception module receives this data, along with additional information from cameras and radars, then publishes data about perceived obstacles in front of the AV.
The prediction module receives the messages published by the perception and localization modules to predict the trajectory of the detected obstacles, and publishes the results to the prediction channel.
The planning module subscribes to messages from all of the previous modules to make driving decisions, e.g., determining the appropriate speed and acceleration.
Finally, the control module converts the trajectory points generated by the planning module into control commands for the chassis, such as steering, throttle, and brake, to ensure the vehicle travels according to the planned trajectory.

\noindent \textbf{Planning module.} 
The ADS's planning module performs three main functions: \textit{route planning}, \textit{behavioral planning}, and \textit{motion planning}~\cite{paden2016survey, gonzalez2015review, schwarting2018planning}.
Given a destination, \textit{route planning} selects a route by choosing a list of lanes and junctions from the map.
This route serves as the reference line for behavioral planning and motion planning.
Behavioral planning is responsible for making high-level driving decisions based on the current driving scenario to interact with pedestrians and other vehicles safely.
For instance, when the AV detects a construction area ahead of its lane, behavioral planning needs to consider both the dynamic behavior of surrounding traffic participants and the road conditions to decide how to bypass it, e.g., by changing lanes.
Lastly, \textit{motion planning} translates high-level decisions into a series of waypoints as part of an executable trajectory, which can be translated into throttles and steering commands by the control module.

\begin{figure}
\centering
\centerline{\includegraphics[width=0.8\linewidth]{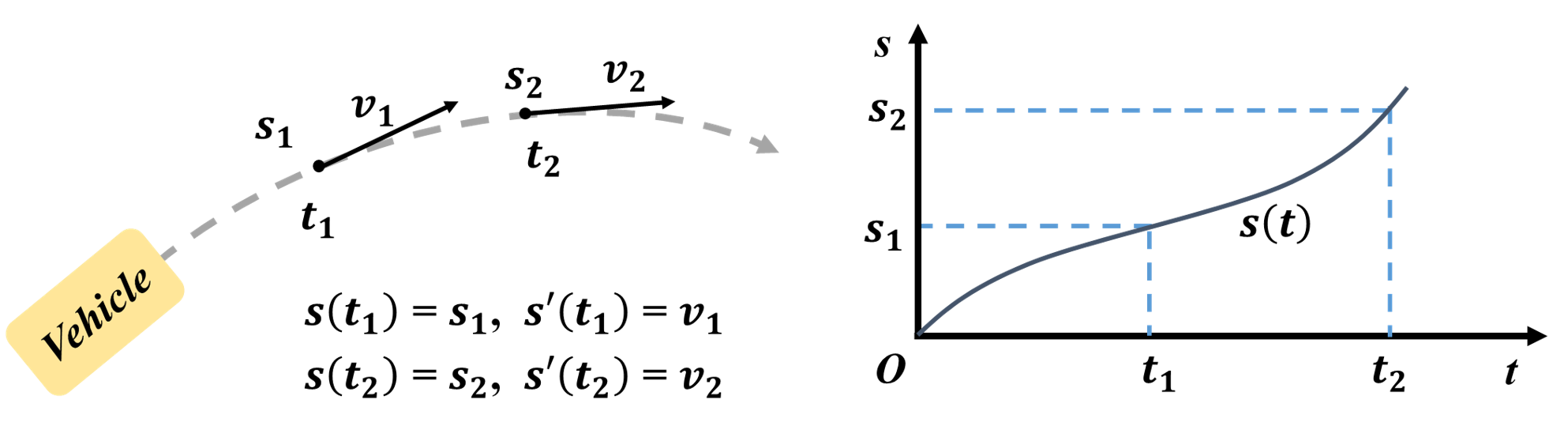}}
\caption{A visual example of an ST graph}
\label{fig:stgraph_eg}
\end{figure}

Behavioral and motion planning are critical tasks of the planning module, translating the path obtained from route planning into a series of waypoints by calculating specific speed and acceleration plans.
This ensures that the AV interacts safely and comfortably with other traffic participants in the current scenario.
Various planning techniques employ distinct approaches to integrate the three essential functions. For example, the lattice planner ~\cite{werling2010optimal}, a graph search-based technique, performs behavioral planning and motion planning implicitly and simultaneously under the guidance of well-designed cost functions. In contrast, the EM planner~\cite{fan2018baidu} performs behavioral planning and motion planning explicitly and step-by-step. In addition, the Frenet frame method is a well-known approach for describing the motion and trajectories of vehicles, which decouples vehicles' lateral and longitudinal motion, corresponding to the lateral and longitudinal control. The longitudinal behavioral and motion planning can be visualized effectively in a Station-Time graph (ST graph), where time is the horizontal axis, the planned longitudinal trajectory distance is the vertical axis and the planned longitudinal trajectory is a curve, as shown in Figure~\ref{fig:stgraph_eg}. Additionally, the curve's gradient represents the longitudinal speed of the vehicle. 

\begin{figure*}[htbp]
\centering  
\subfloat[]{
\includegraphics[width=0.16\textwidth]{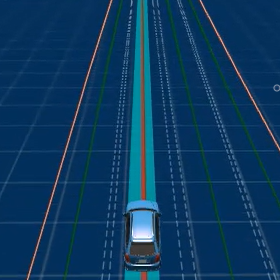}
\label{fig:motivating_example-a}}
\subfloat[]{
\includegraphics[width=0.16\textwidth]{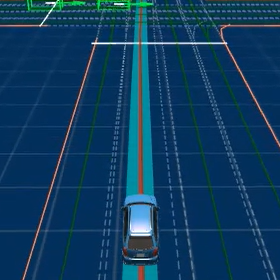}
\label{fig:motivating_example-b}}
\subfloat[]{
\includegraphics[width=0.16\textwidth]{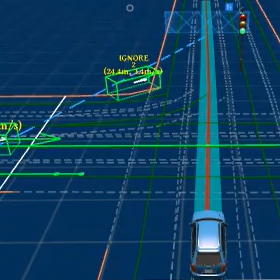}
\label{fig:motivating_example-c}}
\subfloat[]{
\includegraphics[width=0.16\textwidth]{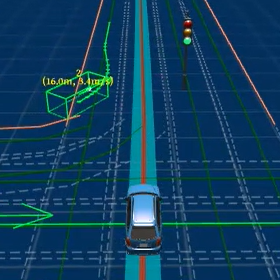}
\label{fig:motivating_example-d}}
\subfloat[]{
\includegraphics[width=0.16\textwidth]{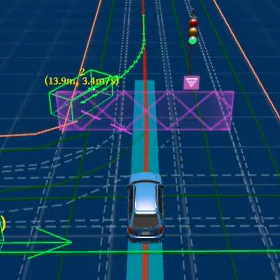}
\label{fig:motivating_example-e}}
\subfloat[]{
\includegraphics[width=0.16\textwidth]{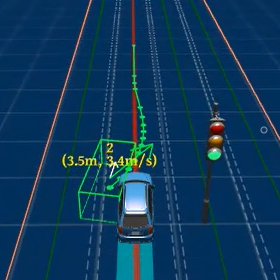}
\label{fig:motivating_example-f}}

\caption{Motivating example: six key scenes from a recording of an AV accident}
\label{fig:motivating_example}
\end{figure*}

\subsection{Motivating Example}
To introduce the concept of accident causality and demonstrate how our framework works, we elaborate with an accident driving recording collected from version 7.0 of Apollo.
As illustrated in Figure~\ref{fig:motivating_example}, we summarize the scenario in an accident driving recording as six critical scenes, the demo video of which can be found on page 3 of `Video Demos' on~\cite{ourweb}. 
Initially, the AV drove alone without encountering any traffic signals (Scene~\ref{fig:motivating_example-a}).
However, it later detected traffic signals and non-player characters (NPCs) as it was approaching an intersection (Scene~\ref{fig:motivating_example-b}).
The AV made an `overtake' decision with respect to NPC 4 and executed it (Scenes~\ref{fig:motivating_example-c}--\ref{fig:motivating_example-d}).
After overtaking NPC 4, it interacted with NPC 2, making a `yield' decision, but still collided with NPC 2 (Scenes \ref{fig:motivating_example-e}--\ref{fig:motivating_example-f}).

\noindent \textbf{Accident-related recording segment.} As we described above, the AV did not detect any NPCs nor was it near any NPCs in Scenes~\ref{fig:motivating_example-a}--\ref{fig:motivating_example-b}.
In contrast to the other four scenes, these first two had no impact on the accident.
Furthermore, during simulation tests, the AV persisted in moving forward \emph{after} a collision, rather than stopping; this behavior, too, had no influence on the accident.
Given that our objective is to perform accident analysis, our framework is designed to automatically identify and exclude such segments that are unrelated to the accident.
The remaining segments are then fed into the causality analysis stage of our framework. For this recording, \framework can significantly reduce its length from 17 seconds to 4 seconds, without removing any critical frames. This reduces the workload of ADS engineers who can then immediately focus on the most important parts of the recording.

\noindent \textbf{Safety-critical frames.} As depicted in Scene~\ref{fig:motivating_example-c}, the AV made an `overtake' decision with respect to NPC 4 near the intersection.
This decision was unsafe because it violates traffic regulations and increases the risk of accidents.
Our framework uses a priori knowledge to label frames containing potential accident risks, such as this one, as safety-critical frames.
Moreover, each of the frames marked as safety-critical (e.g., Scenes \ref{fig:motivating_example-c}--\ref{fig:motivating_example-e}) will be individually inspected by our framework.

\noindent \textbf{Causality analysis.} Our framework can automatically identify causal events in an accident recording, as shown in Table~\ref{tab:motivating_example}.
According to the table, the AV chose to overtake NPC 4 at the intersection due to failing to predict NPC 2's trajectory (Scenes~\ref{fig:motivating_example-c}--\ref{fig:motivating_example-d}).
When the AV finally correctly predicted the trajectory of NPC 2 (Scene~\ref{fig:motivating_example-e}), it was traveling at a speed of 39km/h (approximately 10.83m/s) and was less than 14 meters away from the NPC.
Despite making the appropriate `yield' decision, there was insufficient time and space to carry it out, resulting in a collision.
The causal factors of this accident can be attributed to the incorrect priority and trajectory prediction by the prediction module, the flawed decision made by the planning module, and the vehicle's skidding. 

\section{Framework Design}
\label{sec:framework_design}

As illustrated in Figure~\ref{fig:overview}, our framework consists of two main stages: \emph{accident recording simplification} and \emph{causality analysis}.
In the following sections, we provide a detailed explanation of the two stages of our framework and present an implementation of it for Baidu Apollo 7.0.
Our framework is available online~\cite{ourweb}.

\subsection{Stage \#1: Accident Recording Simplification}

In the first stage, the primary objective is to extract short segments from a long driving recording related to an accident.
The idea is to provide coarse-grained filtering based on the scenario information.
To accomplish this, we partition the driving recording into a series of frames and assign three vectors to each frame to capture information on the current scenario, ranging from the environment to the maneuvers and status of the AV and NPCs.
We use a scenario-based recording segmentation technique to merge the frames, and subsequently, we use specifications derived from the driver handbook and traffic laws of California to identify segments of the recording that are relevant to accidents.

\noindent \textbf{Data Collection.}
A prerequisite for our approach is to collect several accident driving recordings, which requires generating numerous testing scenarios for testing ADSs holistically against different safety oracles.
To satisfy this prerequisite, we adapted the AVUnit framework~\cite{zhou2023specification}, which provides domain-specific languages (DSLs) for specifying testing scenarios and oracles, as well as a fuzzing engine for obtaining effective test cases.
Our adaptation extends the fuzzing engine by adding a recorder that captures the corresponding driving recording for each test case, i.e., each test case is captured in a single recording file.
The set of initial configurations we used in our experiments includes different combinations of starting points, destinations, and NPCs.
To handle the varying routes of all the combinations, the duration of each recording file was set at 60 seconds, which can cover all possible durations of a single test case.
We ran the fuzzing engine for two days, generating 1260 test cases, including 131 accident test cases. The combined length of all these recordings exceeded 21 hours.  
After the termination of the fuzzing algorithm, we selected 110 accident driving recordings based on the output of AVUnit and classified them into three categories (intersection, merging, and rear-end accidents), excluding 21 accident test cases in which the car crash occurred after the AV stopped at its destination or the AV got hit from behind by an NPC.

\begin{table}[!t]
\caption{Results of a causality analysis for the example}
\resizebox{\columnwidth}{!}{%
\begin{tabular}{l|l|l}
\hline
Time & Accident causal events                                                                                             & Details                                                                                                                                                                   \\ \hline
0s   & \begin{tabular}[c]{@{}l@{}}AV keeps safe distance\\ from NPCs\end{tabular}                                         & --                                                                                                                                                                        \\ \hline
0.4s & \begin{tabular}[c]{@{}l@{}}Wrong motion planning;\\ AV skidding sometimes\end{tabular}                             & AV's planning speed is too fast or too slow.                                                                                                                              \\ \hline
0.8s & \begin{tabular}[c]{@{}l@{}}Wrong planning caused \\ by the wrong prediction; \\ AV skidding sometimes\end{tabular} & \begin{tabular}[c]{@{}l@{}}For NPC 2: wrong priority prediction. \\ For NPC 4: improper `overtake' decision. \\ AV's planning speed is too fast or too slow.\end{tabular} \\ \hline
2.6s & \begin{tabular}[c]{@{}l@{}}Wrong motion planning; \\ AV skidding sometimes\end{tabular}                            & AV's planning speed is too fast or too slow.                                                                                                                              \\ \hline
4.3s & Accident!                                                                                                          &                                                                                                                                                                           \\ \hline
\end{tabular}%
}
\label{tab:motivating_example}
\end{table}

\noindent \textbf{Message Alignment and Vectorization.}
In a multi-module ADS, the modules collaborate by asynchronously exchanging and processing messages.
The content of each message varies depending on the module that published it.
To facilitate causality analysis, we select and align messages from the communication channels of the map, localization, perception, prediction, and planning modules, each of which have different publishing frequencies.
We divide the recording into several frames, each of which has a duration of 0.08s (chosen because the localization module has the fastest frequency, publishing messages every 0.08s).
If these channels publish messages within the frame, we hold the messages and align them to the beginning of the frame.
If a channel does not publish any messages within the frame, we copy the last message generated before it to the beginning of the frame.

In the vectorization phase, our primary objective is to extract information related to accidents, which can be associated with the map, perception, prediction, and planning modules in the ADS.
To comprehensively capture information from across these modules' messages, we have designed three feature extraction schemas: one for the map, one for perception and prediction, and one for planning.
Each schema includes factors that impact the AV's planning, or properties that reflect its current planning status.

The map schema contains information on whether the AV is at a junction, crosswalk, or near a stop sign, as well as the color of the perceived traffic signal.
The perception and prediction schema includes four lists of NPCs, indicating which NPCs the AV is approaching, which are in close proximity, and which are predicted as ones to take `caution' of or `ignore'.
The planning schema includes information on the main driving decision the AV currently executes, the operational design domain~(ODD), motion, and whether it is safe according to the responsibility-sensitive safety~(RSS) rules~\cite{DBLP:journals/corr/abs-1708-06374}. It is worth noting that these are fundamental components among industry-level ADSs such as Autoware and Apollo. Specifically, the ODD defines the specific operating conditions and scenarios in which an AV is designed to function safely and effectively. For instance, Autoware's ODDs include `Lane Following', `Lane Change,' and `Pull Out,' among other scenarios, each suggesting the appropriate scene module in Autoware that should be launched to handle the specific driving situation. Similarly, Apollo's ODDs consist of scenarios such as `Lane Change,' `Lane Borrow,' and `Path Assess,' indicating the corresponding decider/optimizer in Apollo that should be activated to make informed driving decisions. To ensure safety and responsible behavior, the planning schema utilizes RSS rules, which are designed to formalize concepts such as dangerous situations, appropriate responses, and the allocation of blame in a mathematically rigorous manner. Our framework converts each frame into three feature vectors based on these three schemas. Each feature vector contains specific semantic properties, with each dimension representing a particular attribute.

For instance, in the feature vector for the map schema, we have four dimensions indicating whether the AV is:
1. Near an intersection (The distance between the AV and the intersection is less than 5 meters);
2. Near a crosswalk (The distance between the AV and the crosswalk is less than 5 meters);
3. Near a stop sign (The distance between the AV and the stop sign is less than 5 meters);
4. Detected traffic signals.
Thus, the vector $\langle False, False, True, None \rangle$ indicates that in the current frame, the AV is approaching a stop sign, not in an area near an intersection or a crosswalk, and not encountering any traffic signals.
In this way, we transform the driving recording into a list of feature vectors while preserving the abstract semantic information of each frame, facilitating subsequent segmenting and pruning.

\noindent \textbf{Segmenting and Pruning.}
After the frame vectorization stage, the framework segments the recording by comparing the similarity of consecutive feature vectors.
The idea is to group together sequential frames with identical feature vectors into a single segment.
For example, if the AV drives on a road segment for 100 uninterrupted frames, then the feature vectors of these 100 frames are the same, and they will be clustered as a single segment based on the static map environment schema.
Our framework generates segmentation plans for each of the three types of vector schemas previously described.
These segmentation plans fuse vectors together using a weighted voting method that determines the optimal clipping point.
For each frame, a general voting function can be defined for any weighted combination of feature vectors.
Let $w_{c}$ denote the weighted value of $c$ feature vectors, and $v_{c}$ denote the vote by the $c$ feature vectors. Let 
\begin{equation}
voting(v_{map}, v_{perc}, v_{pred}) := \sum_{c \in C} w_{c} \times v_{c} \geq \frac{1}{2} \sum_{c \in C} w_{c} 
\label{eq:voting_def}
\end{equation}
\noindent where $C = \lbrace map, perc, pred \rbrace$, which returns $True$ or $False$, indicating (respectively) whether the current vector should be deemed as a clipping point (i.e., last frame of the segment) or not.
The weight of the vote by each category is discussed in Section~\ref{sec:rq_method_design}.

Numerous AV accident reports indicate that most accidents happen in specific contexts, e.g., at intersections, or when there are multiple traffic participants~\cite{favaro2017examining, wang2019exploring, das2020automated}. Armed with this knowledge, our approach creates an overapproximation of relevant frames to narrow down our focus to the most crucial situations. To achieve this, we seek out and discard \emph{irrelevant frames} by analysing static map environments as well as perception and prediction information.
To classify a frame as irrelevant, we consider several factors.
First, we check if the static environment of the frame includes a junction, a crosswalk, or a stop sign.
Next, we verify that the AV is neither approaching nor near any NPC in the frame.
Finally, we ensure that the AV does not predict a `caution' or `ignore' priority for any NPC.
If all of these conditions are met, we classify the frame as an irrelevant frame.
To determine whether to discard a segment $S$, we count the irrelevant frames within it using function $count(S)$, and compute the irrelevant frame ratio $r_m = \frac{count(S)}{len(S)}$.
If $r_m$ is larger than the threshold $th_m$, $S$ will be discarded, otherwise, it will be kept.
We discuss the selection of a particular threshold $th_m$ in Section~\ref{sec:rq_method_design}.

Algorithm~\ref{alg:segmenting_winnowing} summarises the steps of our segmenting and pruning method. 
Specifically, given three categories of feature vectors of an aligned recording, for a feature vector of a frame within it, if the vector is different from its previous one, then we deem that the frame gets one vote by one of the three feature vector categories (Lines 5--10).
After collecting votes from all three categories, we perform voting (Line 11) to decide whether to slice in this frame (Line 12--14), the definition of which is shown in Equation~\ref{eq:voting_def}, and the weight selection of which is discussed in Section~\ref{sec:rq_method_design}. 
We prune the accident-related segments by examining the segments (working backwards) at Lines 17--26.
The last segment is deemed as a part of the accident-related segment (Line 17). 
For other segments, we find the irrelevant map or perception vectors and determine whether to discard them.
For a non-irrelevant segment, we merge it into $S_{a}$ if $S_{a}$ follows it, as shown in Lines 21--25.
We discuss the selection of a threshold value in Section~\ref{sec:rq_method_design}.

\begin{algorithm}[t]
\caption{Segmenting and Pruning}
\label{alg:segmenting_winnowing}
\small
\KwIn{$V$: all the three categories of feature vectors of the original aligned recording before the accident with length $n$;}
\KwOut{$S_{a}$: the reduced accident-related segment; }
$St \gets \varnothing$\; 
$ss \gets 1$\; 
$se \gets 0$\;
\For{$i \gets 2$ to $n$} {
    $v_{map} \gets 0$; $v_{perc} \gets 0$; $v_{pln} \gets 0$ \;
    \For{c in $\lbrace map, perc, pred \rbrace$} {
      \If {$V_{c}[i] \ne V_{c}[i-1]$ or $i == n$} {
        $v_{c} \gets 1$\;
      }
    }
    \tcp{See Section.~\ref{sec:rq_method_design} for voting method}
    \If{voting($v_{map}, v_{perc}, v_{pln}$)} {
      $se \gets i$ \;
      $St.push(V[ss:se])$ \;
      $ss \gets i$ \;
    }
}
$S_{a} \gets St.pop()$ \;
\While{$St.isNotEmpty()$} {
    $S_{curr} \gets St.pop()$ \;
    $r_{m} \gets \frac{count(S_{curr})}{len(S_{curr})}$\;
    \tcp{See Section.~\ref{sec:rq_method_design} for threshold}
    \If{$r_{m} \leq th_{m}$} { 
        \If{$S_{a}$ is succeeded by $S_{curr}$} {
            $S_{a} \gets S_{curr} + S_{a}$\; 
        }
    }
}
\Return $S_{a}$
\end{algorithm}

\subsection{Stage \#2: Causality Analysis}

In the second stage, we automatically analyze the accident-related segments that were generated in the first stage to identify potential causes of the accident.
We utilize automotive safety specifications from California's driver handbook~\cite{ca_drivers_handbook} and traffic laws~\cite{ca_traffic_laws} to identify safety-critical frames that may have contributed to the accident.
Next, we implement a causal analysis tool, \cat, that works by examining speed planning.
For frames that are identified as suspicious, \cat compares their current speed planning and actual trajectory to deduce the causal events of the accident. 
This process enables our framework to effectively identify the causes of the accident and provide valuable insights for future improvements.

\noindent \textbf{Potential Safety-Critical Frame Identification.}
In order to identify safety-critical frames in an accident-related driving recording segment, our framework uses a frame checker that utilizes a priori knowledge, i.e., a list of specifications extracted from background knowledge.
In particular, we examine California's driver handbook~\cite{ca_drivers_handbook}---published by the Department of Motor Vehicles (DMV)---and traffic laws~\cite{ca_traffic_laws}, to obtain a list of specifications for each stage.
These specifications include identifying critical obstacles, improper priority prediction, and driving decision-making. 

In order to ensure compliance with the rules outlined in the driver's handbook~\cite{ca_drivers_handbook}, it is necessary to have a robust specification language that allows us to precisely describe these rules. To this end, we have adopted a specification language based on propositional logic. 
The specification language consists of propositions (based on a set of pre-defined variables), as well as the usual logical connectors.
Before introducing the specifications, we first introduce the pre-defined variables, which can be organized into three categories: state variables, deviation variables, and maneuver variables.

\begin{table}[!t]
	\centering
        \caption{State variables in the specification language}
	\label{tab:state_variables}
	\footnotesize
    \begin{tabular}{c|c|p{2.0in}}
         Variable & Type & Remarks  \\
         \hline
         $x.spd$ & Number & Speed of vehicle x\\ 
         $x.onJct$ & Bool & True if and only if the vehicle x is on a junction \\
         $x.onCswk$ & Bool & True if and only if the vehicle x is on a crosswalk \\
         $x.predTraj$ & Waypoints & vehicle x's predicted trajectory by the EV \\
         $x.isPed$ & Bool & True if and only if the vehicle x is a pedestrian \\
         $x.isBcycl$ & Bool & True if and only if the vehicle x is a bicyclist \\
         $x.bhndEV$ & Bool & True if and only if the vehicle x is in an area behind the EV \\
         $x.blndEV$ & Bool & True if and only if the vehicle x is in the blind area of the EV \\
    \end{tabular}
\end{table}
Firstly, the state variables describe the states of vehicles.
Table~\ref{tab:state_variables} lists a subset of these variables and their usage in describing vehicle properties.
For instance, suppose there is an NPC $npc$ driving near a junction with a speed of 5m/s to the front-left of the AV, then $npc.spd$ is 5m/s, $npc.onJct$ is $True$, and $npc.predTraj$ contains the waypoints in the predicted trajectory of $npc$.
The other variable values of type $bool$ are all $False$.

\begin{table}[!t]
	\centering
        \caption{Deviation variables in the specification language}
	\label{tab:basic_variables}
	\footnotesize
    \begin{tabular}{c|c|p{2.0in}}
         Variable & Type & Remarks  \\
         \hline
         $Th_{err}$ & Number & The threshold of the error of trajectory prediction \\
         $MaxBound$ & Number & The maximum speed limit of a road segment \\
         $MinBound$ & Number & The minimum speed limit of a road segment \\

         $dist(x, y)$ & Number & The distance between two objects $x$ and $y$ \\
         $Err(t)$ & Number & The error of the trajectory prediction $t$ \\
         $CritObst(x)$ & Bool & True if and only if $dist(ev, x) < 3 \times ev.speed$ for an NPC $x$
         \\
    \end{tabular}
\end{table}

Secondly, Table~\ref{tab:basic_variables} summarizes deviation variables to specify various deviation calculations. 
Here, $MaxBound$ and $MinBound$ represent the upper and lower speed limits of the road on which the AV is traveling.
Functions $dist(x, y)$ and $Err(t)$ represent (respectively) the distance between two objects and the error in trajectory prediction.
Additionally, we define the function $CritObst(x)$ to filter out the NPCs that need to be focused on in a given scenario.
The function $CritObst(x)$ outputs $True$ if and only if the distance between object $x$ and the AV is less than three times the current speed of the AV.

\begin{table}[!t]
	\centering
        \caption{Maneuver variables in the specification language}
	\label{tab:ev_maneuver_variables}
	\footnotesize
    \begin{tabular}{c|c|p{2.0in}}
         Variable & Type & Remarks  \\
         \hline
         $PrioIgn(x)$ & Bool & True if and only if the EV predicts NPC $x$ as an "ignore" priority \\
         $DecnIgn(x)$ & Bool & True if and only if the EV makes an "ignore" decision on NPC $x$ \\
         $DecnFlw(x)$ & Bool & True if and only if the EV makes an "follow" decision on NPC $x$ \\
         $DecnYld(x)$ & Bool & True if and only if the EV makes an "yield" decision on NPC $x$ \\
         $DecnOvtk(x)$ & Bool & True if and only if the EV makes an "overtake" decision on NPC $x$ \\
    \end{tabular}
\end{table} 

Finally, the (subset of) maneuver variables presented in Table~\ref{tab:ev_maneuver_variables} reflect the prediction and planning status of the AV.
These variables are directly extracted from prediction and planning messages.
For example, if the AV is closely and cautiously following an NPC $npc$, then $PrioIgn(npc)$ would be $False$, and $DecnFlw(npc)$ would be $True$.
The remaining maneuver variables would be set to $False$.
Here, the AV's priority prediction for an NPC can be roughly divided into three types: `caution' for a critical NPC, `ignore' for an immaterial NPC, and `normal' for the rest.
The AV's driving decision towards an NPC can be summarized as a list of maneuvers, including `ignore', `stop', `follow', `yield', `overtake', `nudge', etc.

With the defined variables, we can now describe the specifications checked by our framework. Specifically, it assesses the correctness of the AV's prioritization, trajectory prediction, driving decisions related to NPCs, and speed planning. For instance, to identify an improper `overtake' decision, we define the specification as:
$ImpropOvtkDecn(x) := (av.OnJct \lor av.OnCswk) \land  DecnOvtk(x)$, which means that if the AV decides to overtake an NPC while near an intersection or on a crosswalk, the `overtake' decision is considered improper. In this case, $x$ refers to perceived objects, such as vehicles, bicycles, or pedestrians. It is important to note that if a specification is satisfied, a vulnerability has been identified. The detailed specifications can be found on our website~\cite{ourweb}.

\noindent \textbf{Causal Events Deduction.}
To identify the causes of accidents from the simplified accident recordings, we design a tool called the Causality Analysis Tool, or \cat for short. 
\cat analyzes frames labeled as safety-critical to determine whether the planning trajectory could intersect with other traffic participants in a way that might cause an accident. If \cat identifies a potential accident scenario, it analyzes the events leading up to that moment and identifies the actions or behaviors that contributed to the scenario. It is worth noting that even if the AV changes its planning in response to a potential accident scenario, incorrect behavior at that moment could waste valuable reaction time and increase the risk of an accident. 

\begin{figure}
\centering
\centerline{\includegraphics[width=1.0\linewidth]{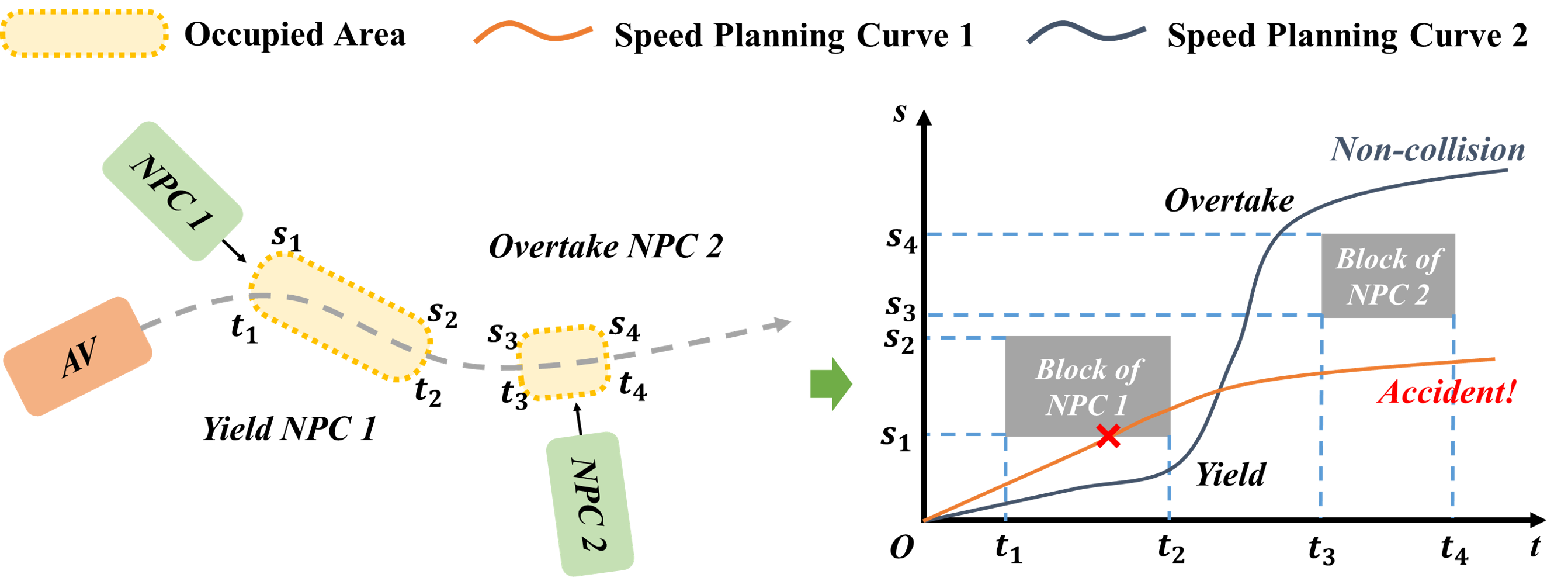}}
\caption{Speed planning based on an ST graph}
\label{fig:st_graph_planning}
\end{figure}

To achieve this, our tool analyzes ST graphs depicting the AV's planning states to discover potential causal events. Based on the Frenet frame method, the ST graph provides a visual way to describe longitudinal behavioral and motion planning. 
Besides directly presenting whether the trajectory plan is collision-free, the ST graph also describes aspects of the AV's driving decisions and speed planning. Specifically, in an ST graph, time is the horizontal axis, the planned longitudinal trajectory distance is the vertical axis and the planned longitudinal trajectory is a curve. Each point on the curve represents a waypoint on the planned trajectory, and the curve's gradient represents the speed. The motion of other traffic participants can be drawn as rectangles that block certain parts of the AV's longitudinal path during a specific time interval. An ideal speed curve intersects with none of these rectangles so that there is no collision between the AV and NPCs. The positional relationship between the speed curve and an obstacle block in the ST graph presents the AV's behavioral planning result for the related traffic participant. If the obstacle block of a traffic participant is above/below the AV's speed curve, the driving decision by the AV is to yield/overtake, as shown in Figure~\ref{fig:st_graph_planning}. Therefore, for achieving collision-free trajectory planning, it is imperative that the vehicle accurately perceives all surrounding NPCs and predicts their future trajectories with high precision. This ensures that there is no overlap between the AV and NPCs at each time step. Fundamentally, this planning process equates to solving a constraint satisfaction problem, where the constraints are defined by the drivable area. In an ideal scenario, precise outputs from the perception and prediction modules would enable the computation to guarantee a collision-free trajectory.

Our tool performs a detailed comparison and analysis of the ST graph from the AV perspective against the ground truth, frame by frame. The idea is that for any given frame in the recording, \cat can reconstruct accurate subsequent trajectories of NPCs using data from the future segments of the recording. This reconstructed trajectory is then treated as the ground truth for assessing the effectiveness of the prediction module. Additionally, we examine the planning module of the tool to verify whether it accurately calculates the necessary constraints for ensuring collision-free trajectory planning for the respective frame.

\begin{figure}
\centering
\centerline{\includegraphics[width=0.9\linewidth]{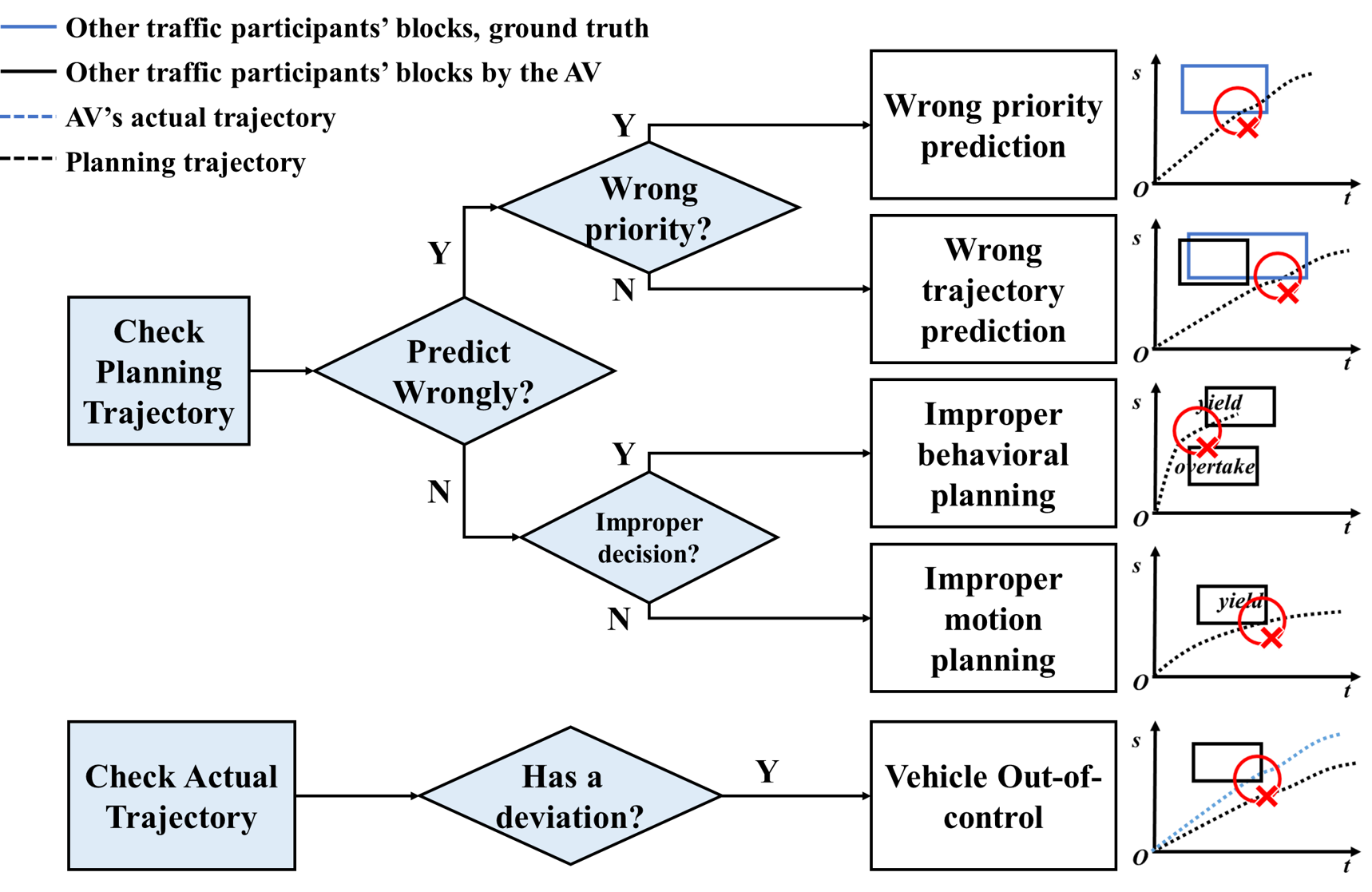}}
\caption{\cat for causality analysis}
\label{fig:cat}
\end{figure}

The analysis process of \cat is shown in Figure~\ref{fig:cat}. \cat firstly checks the priority prediction of the NPC involved in the accident. If the NPC's priority prediction is `ignore', it means that the cause of the collision is wrong priority prediction. This is because AVs do not consider an ignored NPC in the subsequent planning. This omission manifests as a lack of black blocks representing calculated constraints in the ST graph for the NPC, with only the blue blocks indicating the ground truth constraints present.  
If the AV's speed planning curve does not intersect with the obstacle blocks by the AV but intersects with the obstacle block in the ground truth, it means that the cause of the collision is the AV's misunderstanding of the NPC's future action. This situation is characterized by a significant deviation in the ST graph for the NPC, where there is a clear discrepancy between the constraints calculated by the AV and those of the ground truth. 

If the prediction of the NPC made no error, \cat checks the AV's behavioral planning and then the motion planning. 
In the potential safety-critical frame identification step, \framework filters potential improper driving decisions made by the AV. When \cat checks the behavioral planning result, if the speed curve intersects with any obstacle blocks near the risky driving decision, it means that improper behavioral planning is to blame for the accident. For example, the speed curve in the ST graph improperly extends beyond an NPC's block to overtake it. However, in this particular scenario, the AV is unable to find a viable trajectory to avoid a collision with another NPC.
If the speed curve still intersects with other obstacle blocks based on reasonable behavioral planning, it means that improper motion planning caused by risky speed limits is to blame for the accident. In this case, the speed curve in the ST graph demonstrates an insufficient margin relative to the NPC's block, indicating a lack of adequate space to safely avoid the NPC.
If \cat finds that the AV's planning is collision-free, it compares the actual trajectory with the planned trajectory.
If there is a deviation between the two trajectories, we can infer that the AV failed to execute the planning due to being out of control (e.g., due to skidding). 

\noindent \textbf{Generalizability.}
While we have presented \framework in the context of Apollo, the overall approach can be generalized to other ADSs, given that it operates solely on accident recordings and does not require knowledge of the specific internal designs of the systems involved in generating the recordings.
The primary assumption for employing our framework is thus the ability to generate/obtain similar recordings.
Fortunately, modern ADSs typically have multi-module architectures similar to that of Apollo.

We illustrate the generalizability of \framework by applying it to recordings obtained from the Autoware.universe ADS~\cite{autowareuniversegalactic} and the Carla simulator~\cite{dosovitskiy2017carla}.
We systematically examined the semantic structure of message fields required by \framework from various modules, including localization, perception, and planning modules.
In the case of localization messages, there were similarities between the fields in Autoware.universe and Apollo.
Meanwhile, the perception module in Autoware.universe contained tasks related to detecting nearby obstacles and predicting their future trajectories, a functionality akin to the combined roles of perception and prediction modules in Apollo.
Nonetheless, some disparities arose in the message structure. Notably, Autoware.universe lacked an obstacle priority field within perception messages and a behavioral planning field within planning messages. To mitigate these differences in message format, we populated the missing fields with default values. As a result, \framework demonstrated the capability to identify causal events such as wrong trajectory prediction, incorrect speed planning, and instances of vehicles going out of control. However, it was unable to identify causal events related to incorrect priority prediction and erroneous behavioral planning due to the absence of corresponding fields in the message structure (an issue that requires an engineering effort in Autoware.universe to solve).

\section{Evaluation}
\label{sec:evaluation}

\subsection{Research Questions \& Evaluation Metrics}
To evaluate the performance of our framework, we conducted experiments to answer the following research questions:
\begin{itemize}
\item \textbf{RQ1}: Which combination of weights for feature vector categories and which threshold in the ``segmenting and pruning'' phase are the most effective?
\item \textbf{RQ2}: Does \framework effectively simplify accident recording compared to other approaches?
\item \textbf{RQ3}: How many different causal events can the causality analysis of \framework automatically identify?
\item \textbf{RQ4}: To what extent can \framework accurately identify causal events?
\end{itemize}

For RQ1 and RQ2, we evaluated the performance of the simplification methods used in the first stage based on two metrics: the `ratio of reduced frames' and the `recall of critical frames'.
The ratio of reduced frames refers to the length of the removed driving recording over the length of the driving recording before the accident, whereas the recall of critical frames is the number of critical frames in the reduced recording segment over that in the entire recording.
Since the subsequent causality analysis relies on these critical frames, we aimed to preserve them as much as possible.
Therefore, we initially focused on the recall metric of different methods and then considered their ratio of reduced frames.
For RQ3, we assessed the effectiveness of the \framework by analyzing the number of different causal events it could automatically identify based on the simplification of accident recordings. 
For RQ4, we evaluated the accuracy of our framework in identifying causal events resulting from versions of Apollo injected with specific faults.

\subsection{Experiments and Discussion}

\begin{figure}
\centering
\centerline{\includegraphics[width=0.7\linewidth]{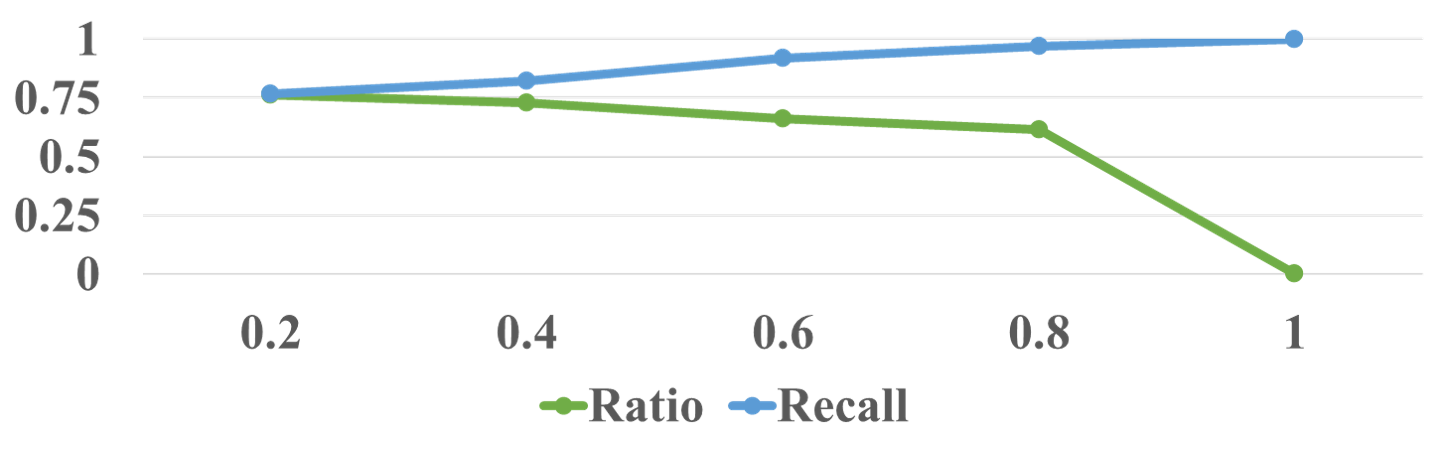}}
\caption{Ratio \emph{(higher is better)} and recall \emph{(higher is better)} of the pruning method under different thresholds}
\label{fig:rq_threshold_ratio_recall}
\end{figure}

\subsubsection{RQ1} \label{sec:rq_method_design}

\begin{table}[t]
\centering
\caption{Ratio \emph{(higher is better)} and recall \emph{(higher is better)} of different combinations of voting methods}
\resizebox{\columnwidth}{!}{%
\begin{tabular}{c|c|c|c|c|c|c|c}
\hline
\begin{tabular}[c]{@{}c@{}}Weight Ratio\\ (map:perc:pln)\end{tabular} & 1:1:0   & 1:0:1   & 0:1:1   & \textbf{1:1:1}   & 2:1:1   & 1:2:1   & \textbf{1:1:2}   \\ \hline
Ratio                                                                 & 74.64\% & 96.43\% & 74.64\% & \textbf{50.03\%} & 60.26\% & 74.64\% & \textbf{62.23\%} \\ \hline
Recall                                                                & 79.62\% & 11:06\% & 79.62\% & \textbf{93.01\%} & 89.19\% & 79.62\% & \textbf{94.41\%} \\ \hline
\end{tabular}%
}
\label{tab:rq_segmentation_method}
\end{table}

Different segmenting methods lead to different segmentations of the recording, which can affect the efficacy of test reduction and the final analysis. This is due to the varying contributions of features in depicting a driving scenario. Additionally, using the same contribution for all features can result in many short clips and a lower reduction ratio of original recordings. To design a coarse-grained test reduction method, we evaluated the effectiveness of various combinations of weights assigned to categories of feature vectors and the threshold for determining accident recording segments in RQ1. This method aims to identify and remove non-accident segments to reduce the overall size of the recording for analysis.

We first evaluated the performance of different settings of the voting method for frame segmenting when the threshold value was set as 0.8 for pruning and present the results in Table~\ref{tab:rq_segmentation_method}. We focused on the recall of critical frames, as this factor can significantly impact the causality analysis conducted by our a priori frame checker and \cat. The results indicated that the voting method with a weight ratio of $map:perc:pln = 1:1:2$ (i.e., the method adopted by our framework) achieved the best total recall rate of 94.41\% across all frame segmenting methods. This method also had a reduced frame ratio of 62.23\%, signifying its effectiveness in removing non-accident recording segments from the analysis. It is also worth noting that the voting method with a weight ratio of $map:perc:pln = 1:1:1$ achieved a similar recall rate (93.01\%) compared to our method (94.41\%) while having the lowest ratio of reduced frames among all the weight combinations. However, we observed that the segments generated by this weight combination were fewer in number and larger in length than those created by our segmenting method, leading to fewer segments being discarded in the recording pruning stage. As a result, insufficient recording pruning allowed this method to maintain a promising recall, but it does not necessarily imply that this is an effective segmenting method. The optimal balance between recall and pruning efficiency is crucial for an effective segmenting method, and our method with the weight ratio $map:perc:pln = 1:1:2$ has demonstrated better overall performance in capturing critical frames and pruning irrelevant ones.

In order to determine the optimal threshold for our segment pruning method, we conducted a series of experiments, adjusting the threshold for identifying accident-related segments in increments of 0.2, starting from 0.2. We focused on the same two metrics: recall and ratio. The results presented in Figure~\ref{fig:rq_threshold_ratio_recall} reveal that as the threshold value increases, recall progressively improves. When the threshold value exceeds 0.4, recall consistently remains above 80\%. Simultaneously, the ratio gradually decreases as the threshold value rises. From a threshold of 0.2 to 0.8, the ratio experiences minimal change and maintains a level above 60\%. However, when the threshold increases from 0.8 to 1, the ratio experiences a substantial decrease compared to previous levels. Based on these findings, we concluded that a threshold of 0.8 is optimal, as it strikes a balance between high record reduction performance and the retention of a sufficient number of safety-critical frames.

\subsubsection{RQ2}

\begin{table}[!t]
\centering
\caption{Ratio \emph{(higher is better)} and recall \emph{(higher is better)} of different recording segmenting methods}
\resizebox{\columnwidth}{!}{%
\begin{tabular}{c|c|c|c|c|c|c}
\hline
       & \framework    & STRaP   & Length: 4s & Length: 8s & Length: 12s & Length: 16s \\ \hline
Ratio  & 62.23\% & 60.57\% & 76.74\%    & 54.40\%    & 32.64\%     & 16.29\%     \\ \hline
Recall & 94.41\% & 30.81\% & 72.26\%    & 82.92\%    & 86.85\%     & 91.35\%     \\ \hline
\end{tabular}%
}
\label{tab:rq_other_methods}
\end{table}

For RQ2, our objective is to compare our accident recording simplification method with a variety of alternative fixed-length recording pruning methods and the STRaP framework \cite{deng2022scenario}, an AV recording simplification method.
We set the lengths at 4, 8, 12, and 16 seconds before the accident, considering that the remaining segment length of our approach is approximately between 4s and 16s. The results of our experiment are displayed in Table~\ref{tab:rq_other_methods}. The rows represent the evaluation metrics of different segmenting and pruning methods, while the columns indicate the various accident categories included in the experiments. A comparison with fixed-length segmenting methods reveals that it is not feasible to establish a fixed remaining length that effectively balances a substantial reduction ratio with a high critical frame recall. Upon further examination of the accident-related segment lengths, we believe that the primary reason for this outcome is the variability in the duration of interaction between the AV and the NPC involved in different accidents. This observation also highlights the utility and generalizability of our approach, which can adapt to a wide range of cases. 

As our segmenting and pruning method shares similar goals with the concept of test reduction and prioritization, we further compared our accident recording simplification method with STRaP, which scales redundant segments with similar contents down to a given length to reduce the length of a recording. As shown in RQ1, \framework's ratio of reduced frames is 62.23\% on average. Therefore, we restricted the retained recording length in STRaP as 40\% of the number of frames in the original segment to ensure a similar reduced frame ratio, i.e., a ratio rate of about 60\%. In our experiment, STRaP achieved a total reduced frame ratio rate of 39.43\% and a recall rate of 30.81\%. The reason is that the STRaP framework, while effective in its intended purpose, modifies the content of the original recordings in such a way that distorts the temporal relationships between events and their true durations. This alteration of the original recordings makes STRaP unsuitable for causality analysis.

\begin{table}[!t]
\caption{The number of causal events over different accident types}
\resizebox{\columnwidth}{!}{%
\begin{tabular}{c|c|c|c|c|c}
\hline
               & \begin{tabular}[c]{@{}c@{}}Wrong\\ Priority\\ Prediction\end{tabular} & \begin{tabular}[c]{@{}c@{}}Wrong\\ Trajectory\\ Prediction\end{tabular} & \begin{tabular}[c]{@{}c@{}}Wrong\\ Behavioral\\ Planning\end{tabular} & \begin{tabular}[c]{@{}c@{}}Wrong\\ Motion\\ Planning\end{tabular} & \begin{tabular}[c]{@{}c@{}}Vehicle\\ Out-of-control\end{tabular} \\ \hline
\textbf{Total} & \textbf{26}                                                           & \textbf{51}                                                             & \textbf{17}                                                           & \textbf{67}                                                       & \textbf{103}                                                     \\ \hline
Intersection   & 0                                                                     & 0                                                                       & 6                                                                     & 27                                                                & 39                                                               \\ \hline
Merging        & 20                                                                    & 27                                                                      & 4                                                                     & 23                                                                & 30                                                               \\ \hline
Tailgating     & 6                                                                     & 24                                                                      & 7                                                                     & 17                                                                & 34                                                               \\ \hline
\end{tabular}%
}
\label{tab:effectiveness}
\end{table}

\subsubsection{RQ3}
In RQ3, our objective is to determine \framework's performance on the accident recordings collected for the original ADS. To achieve this, we conducted a comprehensive evaluation by applying our framework to a dataset comprising 110 accident recordings, all generated by an AV testing engine \cite{zhou2023specification}. This dataset encompassed a variety of accident scenarios, including 43 intersection accidents, 31 merging accidents, and 36 rear-end accidents. \framework successfully identified the causal events for 103 of these accident recordings. However, our study found that \framework was unable to detect any significant causal events in 7 accident recordings. Upon further examination, we discovered that these accidents merely involved minor scratches between the AV and the NPC, without any severe impacts taking place. This issue can be attributed to the limitations of computational precision, which can be perceived as an engineering challenge arising from the complexities of accurately processing distances.

For the remaining 103 accidents, we conducted a manual verification process. This entailed revisiting all the causality analysis reports by replaying the accident recordings and validating the causal events identified by \framework. In particular, we conducted a systematic examination of all causal events identified by our framework and present the specific numbers for each accident type in Table~\ref{tab:effectiveness}. These results indicate that \framework can effectively identify multiple causal events in various accidents, utilizing each causal event defined by CAT. Through \framework, the events of wrong trajectory prediction were primarily found in merging and rear-end type accidents, while wrong speed planning events occurred more frequently in intersection and merging accidents. It is important to note that all the accidents in our dataset occurred in rainy or snowy weather conditions, which explains the “vehicle out-of-control” event appearing in all 103 accidents.

\begin{table*}[!t]
\centering
\caption{The eight types of faults injected into the customized ADS}
\label{tab:rq_faultinjections}
\resizebox{\textwidth}{!}{%
\begin{tabular}{c|c|l}
\hline
Fault Type & Location                                      & Description                                                                                                           \\ \hline
F1         & $\mathtt{AssignIgnoreLevel()@obstacle\_prioritizer.cc}$  & Assign `ignore' priority to all the detected NPCs by default.                                                           \\ \hline
F2         & $\mathtt{PredictObstacle()@predictor\_manager.cc}$       & Assign improper trajectory prediction models to NPCs to get erroneous trajectory prediction.                          \\ \hline
F3         & $\mathtt{MakeStaticObstacleDecision()@path\_decider.cc}$ & Make `ignore' decisions to all the static NPCs near the AV's planned trajectory.                                        \\ \hline
F4         & $\mathtt{MakeObjectDecision()@speed\_decider.cc}$        & Make `follow' decisions to any NPCs in front of the AV which tend to stop, instead of `stop' decisions or changing lanes.  \\ \hline
F5         & $\mathtt{MakeObjectDecision()@speed\_decider.cc}$        & Make `ignore' decisions to an NPC ahead of the AV, if the AV is not following or keeping distance from it.              \\ \hline
F6         & $\mathtt{MakeObjectDecision()@speed\_decider.cc}$        & Make `yield' decisions to a high-speed NPC accelerating ahead of the AV, which leads to AV's low speed in a fast lane. \\ \hline
F7         & $\mathtt{MakeObjectDecision()@speed\_decider.cc}$        & Make `overtake' decisions to any NPC if it is near the AV.                                                              \\ \hline
F8         & $\mathtt{GetSpeedLimits()@speed\_limit\_decider.cc}$     & Keep a high speed even being close to NPCs.                                                                           \\ \hline
\end{tabular}%
}
\end{table*}

\begin{table}[!t]
\caption{Precision \emph{(higher is better)}, Recall \emph{(higher is better)}, 
 and Accuracy \emph{(higher is better)} of causal events over accidents with different fault injections}
\resizebox{\columnwidth}{!}{%
\begin{tabular}{c|ccc|ccccccc}
\hline
Location       & \multicolumn{3}{c|}{Prediction Module}                                      & \multicolumn{7}{c}{Planning Module}                                                                                                                                                                        \\ \hline
Fault Types    & \multicolumn{1}{c|}{F1}     & \multicolumn{1}{c|}{F2}     & Total           & \multicolumn{1}{c|}{F3}     & \multicolumn{1}{c|}{F4}             & \multicolumn{1}{c|}{F5}     & \multicolumn{1}{c|}{F6}     & \multicolumn{1}{c|}{F7}     & \multicolumn{1}{c|}{F8}     & Total          \\ \hline
Numbers        & \multicolumn{1}{c|}{155}    & \multicolumn{1}{c|}{126}    & 281             & \multicolumn{1}{c|}{132}    & \multicolumn{1}{c|}{146}            & \multicolumn{1}{c|}{202}    & \multicolumn{1}{c|}{166}    & \multicolumn{1}{c|}{145}    & \multicolumn{1}{c|}{134}    & 925            \\ \hline
Precision (\%) & \multicolumn{1}{c|}{100.00} & \multicolumn{1}{c|}{100.00} & 100.00          & \multicolumn{1}{c|}{100.00} & \multicolumn{1}{c|}{100.00}         & \multicolumn{1}{c|}{100.00} & \multicolumn{1}{c|}{100.00} & \multicolumn{1}{c|}{100.00} & \multicolumn{1}{c|}{100.00} & 100.00         \\ \hline
Recall (\%)    & \multicolumn{1}{c|}{100.00} & \multicolumn{1}{c|}{90.00}  & 95.97           & \multicolumn{1}{c|}{87.61}  & \multicolumn{1}{c|}{73.98}          & \multicolumn{1}{c|}{89.70}  & \multicolumn{1}{c|}{86.72}  & \multicolumn{1}{c|}{77.19}  & \multicolumn{1}{c|}{77.19}  & 82.56          \\ \hline
Accuracy (\%)  & \multicolumn{1}{c|}{100.00} & \multicolumn{1}{c|}{92.06}  & \textbf{96.44}  & \multicolumn{1}{c|}{89.39}  & \multicolumn{1}{c|}{\textbf{78.08}} & \multicolumn{1}{c|}{91.58}  & \multicolumn{1}{c|}{89.76}  & \multicolumn{1}{c|}{82.07}  & \multicolumn{1}{c|}{80.60}  & \textbf{85.73} \\ \hline
\end{tabular}%
}
\label{tab:recall_precision}
\end{table}

\subsubsection{RQ4}

In RQ4, we sought to assess the accuracy of our framework in identifying the causes of accidents. To achieve this, we injected eight distinct fault types, as detailed in Table~\ref{tab:rq_faultinjections}, into the ADS. Specifically, F1 can cause an accident due to wrong priority prediction causal events, while F2 causes accidents based on wrong trajectory prediction. Conversely, F3 through F7 are designed to cause accidents due to improper behavioral planning. Finally, F8 is identified as the trigger for a causal event related to improper motion planning. For each fault type, we ran the testing engine~\cite{zhou2023specification} for approximately one day and recorded the resulting accidents. It is imperative to highlight our efforts to ensure the complexity of each recorded test case. We accomplished this by implementing varying extended routes and incorporating multiple NPCs of diverse types. Furthermore, we standardized the duration of each recording file to 120 seconds. In total, we amassed a dataset comprising 1206 accident recordings. 

Subsequently, we applied our framework to analyze these accident recordings, documenting the accident causal events and their respective time frames.
In this experiment, if a causal event's duration significantly surpassed those of other causal events, we deemed it to be the `main' cause of the given accident.
For example, in the case of fault F2, if the injected fault takes effect, it should persist for a sufficient duration to accumulate a noticeable trajectory prediction error, which is crucial for causing accidents. Consequently, the associated causal event, namely, `wrong trajectory prediction', would be identified in the recording files as the main cause of this fault. If our framework correctly identifies the functions in line with the injected faults, we conclude that our framework accurately determines the cause of the accidents.

As shown in Table~\ref{tab:recall_precision}, \framework performs well, accurately identifying causal events in 1064 out of the accident recordings, with a precision of 100.00\% for both the prediction and planning modules. This indicates that, for a specific type of fault, our framework can both precisely identify the causal events within the recording and distinguish recordings that do not include these causal events. Furthermore, this is complemented by a recall rate of 95.97\% and an accuracy of 96.44\% in the prediction module, along with a recall rate of 82.56\% and an accuracy of 85.73\% in the planning module.

Nevertheless, our investigation uncovered a limitation, as \framework failed to detect the causal events in 142 accident recordings. Upon a more in-depth examination, we discovered that when faults are injected into the planning module, two or three closely interrelated causal events often occur simultaneously. For instance, in 15 accidents linked to fault F7, an additional causal event surfaced: the vehicle going out of control. This event was attributable to the elevated speed requirement of the `overtake' decision, particularly evident during inclement weather conditions like rain or snow. 
We observed that \framework successfully identified interrelated causal events in 105 out of the 142 accidents.

\subsubsection{Threats to Validity}

We acknowledge certain limitations and threats to the validity of our evaluation. 
While our approach has been implemented for two distinct platforms---Apollo, simulated with the SVL Simulator, and Autoware.universe, simulated with Carla---our evaluation is exclusively focused on the Apollo ADS. The reason is that there is currently no suitable fuzzing engine implemented for Autoware.universe. This absence presents a challenge in acquiring sufficient accident recordings for a comprehensive evaluation of our approach on the platform.
Second, during testing, we observed that the AV primarily considered NPCs in front of it when planning driving behavior. When an NPC hits the AV from behind, \framework may not yield effective analysis results. This issue could be addressed by incorporating more intelligent NPC behavior configurations in the simulator, which would better emulate interactions between real-world vehicles. Furthermore, it is generally accepted that the rear vehicle should bear more responsibility in a rear-end accident, a principle that is also practiced in many jurisdictions \cite{rear_end_fault_compensation}.
Third, it is imperative to acknowledge that the faults injected in RQ4 do not reflect the real-world faults in the ADSs. However, the resulting accidents from these injected faults are similar to those caused by real-world faults in ADSs, lending credence to our framework's ability to accurately identify causal factors of accidents. Moreover, the inherent complexity of ADSs, attributed to their reliance on logic-based code, external dependency libraries, and machine learning-based models across various modules, contributes to a significant challenge for repairing. As reported in a study \cite{10.1145/3377811.3380397}, more than half of the AV faults originate from incorrect algorithmic implementations or configurations, often involving extensive code segments exceeding 20 lines. Consequently, while our framework can interpret accident recordings and pinpoint potential causes, it should not be considered a panacea for repairing the underlying bugs in ADS systems.

\section{Related Work}
\label{sec:related_work}

System-level testing for AVs is designed to evaluate the performance of the entire ADS, as opposed to module-level testing, which focuses on individual modules or specific functionalities.
This comprehensive evaluation is achieved through the use of scenario-based test cases and test oracles.
Current research in system-level testing primarily focuses on generating corner cases and error-prone driving scenarios.
There are two main categories of scenario sources: real-world data, and testing frameworks.

One category of work generates scenarios derived from scenarios observed in the real world, emphasizing the similarity between the generated scenarios and real-world ones \cite{paardekooper2019automatic, nitsche2017pre}.
Zhang et al.~\cite{zhang2014roadview} proposed a method based on 3D scene reconstruction, which uses images collected by the in-vehicle camera to recreate scenarios as test cases.
Gambi et al.~\cite{dreossi2019verifai} proposed AC3R, which extracts information from collision reports and constructs new test scenarios using simulation methods.
DEEPCRASHTEST~\cite{bashetty2020deepcrashtest} recreates accident scenarios based on accident videos.
Fremont et al.~\cite{fremont2020formal} combined formal verification with clustering algorithms to select usable test scenarios.
There is also an approach \cite{roesener2016scenario} that evaluates the performance of the ADSs by comparing them with that of human drivers according to features extracted from real-world scenarios.

Another category of work generates scenarios by using a (domain-specific) testing framework.
Two widely-adopted methodologies are search-based or sampling-based methods \cite{dreossi2019compositional, abdessalem2018testing_mos, riccio2020model, zohdinasab2021deephyperion, abdessalem2018testing, luo2021targeting, hauer2019did, beglerovic2017testing}.
Search-based methods, or fuzzing, typically search the parameter space for specific parameter values to achieve a certain testing goal.
To guarantee the efficiency of the heuristic search method adopted, e.g., genetic algorithms, a well-defined fitness function is required.
Althoff et al.~\cite{althoff2017commonroad} defined a calculating metric, the drivable area, to quantify the search of solution space, and combined reachability analysis with optimization techniques to obtain test scenarios.
Li et al.~\cite{li2020av} proposed AVFuzzer, which uses safety potential, the distance between the ego vehicle and other traffic participants, as the fitness function for a genetic algorithm-based fuzzer to find scenarios that could lead to collisions.
Combining program analysis techniques and evolutionary algorithm-based fuzzing, PlanFuzz~\cite{ndss:2022:ziwen:planfuzz} defines behavioral planning vulnerability distance as the guidance for the generation of test scenarios that would cause the autonomous vehicle to stop under safe conditions.
Sun et al.~\cite{sun2022lawbreaker} defined a metric for quantifying the degree to which autonomous vehicles violate traffic rules in a driving scenario, guiding their fuzzer to generate test cases that violate traffic regulations.
Sampling-based methods sample from a naturalistic scenario distribution to generate test cases.
A series of works~\cite{zhao2016accelerated, huang2017accelerated, zhao2017accelerated, wang2021combining} has studied sampling in different driving scenarios based on importance sampling~\cite{tokdar2010importance}.
Batsch et al.~\cite{birkemeyer2022feature} built a Gaussian process classification model to estimate the safety of a scenario probabilistically, with the training data sampled from simulation-generated traffic congestion scenarios.
NADE~\cite{feng2021intelligent} collects driving scenarios from real-world data and samples to generate realistic and safety-critical scenarios.

The aforementioned works primarily concentrate on evaluating the performance of ADSs comprehensively and identifying new vulnerabilities. However, their focus lies in determining whether the ADSs fail to meet the test oracles, rather than understanding the underlying reasons for these failures. Our method is driven by the goal of analyzing the actual cause of safety violations, such as collisions, by concentrating on the testing process itself. 

In recent years, causality has become a widely-adopted methodology to analyze complex systems.
Forney et al.~\cite{ibrahim2020actual} proposed an interactive platform for fault diagnosis and forensic investigation in fields such as airplane accidents.
Bareinboim et al.~\cite{bareinboim2016causal} proposed a causal inference-based method to solve data fusion problems in the context of big data. 
Biebl et al.~\cite{DBLP:conf/hci/BieblKUTRLPB21} presented a causal model to predict accident risks in an intersection for drivers with impairments.
In addition to works focusing on AI~\cite{zhang2017transfer, correa2017causal, correa2018generalized, bareinboim2015recovering}, some works have applied causality to the security analysis of CPSs~\cite{moradi2020actor, nigam_kim_mason_talcott_2022, JADIDI2022103741, DBLP:conf/nfm/IbrahimKPHK19, DBLP:journals/corr/abs-2005-03294}.
Zhang et al.~\cite{zhang2020tracing} monitored, inspected, and located anomalies in industrial control systems using a causal model based on maximum information coefficient and transfer entropy.
Poskitt et al.~\cite{Poskitt-et_al23a} proposed a causality-guided fuzzing method that identifies and generalizes the causality of events in testing to find new test cases with different causal relationships. 
Our method is designed to employ causality analysis on autonomous driving accident records to facilitate deeper fault analysis and uncover the underlying causes of accidents. By examining the causal factors that contribute to accidents, we can better understand the limitations and vulnerabilities of autonomous driving systems. This, in turn, allows engineers to make more targeted improvements, enhance safety, and reduce the likelihood of similar accidents occurring in the future.  

\section{Conclusion}
\label{sec:conclusion}
We presented \framework, an automated framework for determining the causal events in AV accidents. We successfully implemented it in both Apollo and Autoware.universe and evaluated our framework using 110 accident driving recordings from the Baidu Apollo ADS, successfully identifying causal events in 103 of them.
After analyzing 1206 accident recordings collected from ADSs injected with specific faults, we further showed that it identifies causal events correctly.

In future work, we are interested in developing automatic program repair techniques for ADSs, leveraging the results of causality analyses from accidents. 
By incorporating these advancements, we hope to create a comprehensive framework that can contribute significantly to the safety and reliability of AVs in real-world scenarios.

\section*{Acknowledgment}
We extend our sincere appreciation to the three anonymous referees from ICSE for their invaluable insights and constructive feedback. 
This research was jointly sponsored by Shanghai Sailing Program under Grants No.~23YF1427500, the NSFC Program under Grants No.~62302304, ShanghaiTech Startup Funding, and the Ministry of Education, Singapore under its Academic Research Fund Tier 3 (Award ID: MOET32020-0004). Any opinions, findings and conclusions or recommendations expressed in this material are those of the author(s) and do not reflect the views of the Ministry of Education, Singapore.

\bibliographystyle{ACM-Reference-Format}

\bibliography{reference}


\begin{thebibliography}{64}


\ifx \showCODEN    \undefined \def \showCODEN     #1{\unskip}     \fi
\ifx \showDOI      \undefined \def \showDOI       #1{#1}\fi
\ifx \showISBNx    \undefined \def \showISBNx     #1{\unskip}     \fi
\ifx \showISBNxiii \undefined \def \showISBNxiii  #1{\unskip}     \fi
\ifx \showISSN     \undefined \def \showISSN      #1{\unskip}     \fi
\ifx \showLCCN     \undefined \def \showLCCN      #1{\unskip}     \fi
\ifx \shownote     \undefined \def \shownote      #1{#1}          \fi
\ifx \showarticletitle \undefined \def \showarticletitle #1{#1}   \fi
\ifx \showURL      \undefined \def \showURL       {\relax}        \fi
\providecommand\bibfield[2]{#2}
\providecommand\bibinfo[2]{#2}
\providecommand\natexlab[1]{#1}
\providecommand\showeprint[2][]{arXiv:#2}

\bibitem[fat(2019)]%
        {fatal_tesla_crash}
 \bibinfo{year}{2019}\natexlab{}.
\newblock \bibinfo{title}{Fatal Tesla Crash Exposes Gap In Automaker's Use Of Car Data}.
\newblock \bibinfo{howpublished}{\url{https://www.forbes.com/sites/alanohnsman/2018/04/16/tesla-autopilot-fatal-crash-waze-hazard-alerts/?sh=229d9e685572}}.
\newblock
\newblock
\shownote{Online; accessed December 2023}.


\bibitem[apo(2021)]%
        {apollo70}
 \bibinfo{year}{2021}\natexlab{}.
\newblock \bibinfo{title}{Apollo 7.0}.
\newblock \bibinfo{howpublished}{\url{https://github.com/ApolloAuto/apollo/releases/tag/v7.0.0}}.
\newblock
\newblock
\shownote{Online; accessed April 2023}.


\bibitem[aut(2021)]%
        {autowareuniversegalactic}
 \bibinfo{year}{2021}\natexlab{}.
\newblock \bibinfo{title}{Autoware.universe galactic}.
\newblock \bibinfo{howpublished}{\url{https://github.com/autowarefoundation/autoware.universe/tree/galactic}}.
\newblock
\newblock
\shownote{Online; accessed November 2023}.


\bibitem[bay(2022)]%
        {bay_city_news}
 \bibinfo{year}{2022}\natexlab{}.
\newblock \bibinfo{title}{{Fatal crash on SB I-680 onramp in San Jose}}.
\newblock \bibinfo{howpublished}{\url{https://www.kron4.com/news/bay-area/fatal-crash-on-sb-i-680-onramp-in-san-jose/}}.
\newblock
\newblock
\shownote{Online; accessed December 2023}.


\bibitem[rea(2022)]%
        {rear_end_fault_compensation}
 \bibinfo{year}{2022}\natexlab{}.
\newblock \bibinfo{title}{Rear-End Collisions: Fault \& Compensation}.
\newblock \bibinfo{howpublished}{\url{https://www.forbes.com/advisor/legal/auto-accident/rear-end-collision/\#establishing_fault_for_rear_end_accidents_section}}.
\newblock
\newblock
\shownote{Online; accessed December 2023}.


\bibitem[our(2023)]%
        {ourweb}
 \bibinfo{year}{2023}\natexlab{}.
\newblock \bibinfo{title}{ACAV2023}.
\newblock \bibinfo{howpublished}{\url{https://acav2023.github.io}}.
\newblock
\newblock
\shownote{Online; accessed December 2023}.


\bibitem[Abdessalem et~al\mbox{.}(2018a)]%
        {abdessalem2018testing}
\bibfield{author}{\bibinfo{person}{Raja~Ben Abdessalem}, \bibinfo{person}{Shiva Nejati}, \bibinfo{person}{Lionel~C Briand}, {and} \bibinfo{person}{Thomas Stifter}.} \bibinfo{year}{2018}\natexlab{a}.
\newblock \showarticletitle{Testing vision-based control systems using learnable evolutionary algorithms}. In \bibinfo{booktitle}{\emph{Proceedings of the 40th International Conference on Software Engineering}}. \bibinfo{pages}{1016--1026}.
\newblock


\bibitem[Abdessalem et~al\mbox{.}(2018b)]%
        {abdessalem2018testing_mos}
\bibfield{author}{\bibinfo{person}{Raja~Ben Abdessalem}, \bibinfo{person}{Annibale Panichella}, \bibinfo{person}{Shiva Nejati}, \bibinfo{person}{Lionel~C Briand}, {and} \bibinfo{person}{Thomas Stifter}.} \bibinfo{year}{2018}\natexlab{b}.
\newblock \showarticletitle{Testing autonomous cars for feature interaction failures using many-objective search}. In \bibinfo{booktitle}{\emph{Proceedings of the 33rd ACM/IEEE International Conference on Automated Software Engineering}}. \bibinfo{pages}{143--154}.
\newblock


\bibitem[Althoff et~al\mbox{.}(2017)]%
        {althoff2017commonroad}
\bibfield{author}{\bibinfo{person}{Matthias Althoff}, \bibinfo{person}{Markus Koschi}, {and} \bibinfo{person}{Stefanie Manzinger}.} \bibinfo{year}{2017}\natexlab{}.
\newblock \showarticletitle{CommonRoad: Composable benchmarks for motion planning on roads}. In \bibinfo{booktitle}{\emph{2017 IEEE Intelligent Vehicles Symposium (IV)}}. IEEE, \bibinfo{pages}{719--726}.
\newblock


\bibitem[Bareinboim and Pearl(2016)]%
        {bareinboim2016causal}
\bibfield{author}{\bibinfo{person}{Elias Bareinboim} {and} \bibinfo{person}{Judea Pearl}.} \bibinfo{year}{2016}\natexlab{}.
\newblock \showarticletitle{Causal inference and the data-fusion problem}.
\newblock \bibinfo{journal}{\emph{Proceedings of the National Academy of Sciences}} \bibinfo{volume}{113}, \bibinfo{number}{27} (\bibinfo{year}{2016}), \bibinfo{pages}{7345--7352}.
\newblock


\bibitem[Bareinboim and Tian(2015)]%
        {bareinboim2015recovering}
\bibfield{author}{\bibinfo{person}{Elias Bareinboim} {and} \bibinfo{person}{Jin Tian}.} \bibinfo{year}{2015}\natexlab{}.
\newblock \showarticletitle{Recovering causal effects from selection bias}. In \bibinfo{booktitle}{\emph{Proceedings of the AAAI Conference on Artificial Intelligence}}, Vol.~\bibinfo{volume}{29}.
\newblock


\bibitem[Bashetty et~al\mbox{.}(2020)]%
        {bashetty2020deepcrashtest}
\bibfield{author}{\bibinfo{person}{Sai~Krishna Bashetty}, \bibinfo{person}{Heni~Ben Amor}, {and} \bibinfo{person}{Georgios Fainekos}.} \bibinfo{year}{2020}\natexlab{}.
\newblock \showarticletitle{Deepcrashtest: Turning dashcam videos into virtual crash tests for automated driving systems}. In \bibinfo{booktitle}{\emph{2020 IEEE International Conference on Robotics and Automation (ICRA)}}. IEEE, \bibinfo{pages}{11353--11360}.
\newblock


\bibitem[Beglerovic et~al\mbox{.}(2017)]%
        {beglerovic2017testing}
\bibfield{author}{\bibinfo{person}{Halil Beglerovic}, \bibinfo{person}{Michael Stolz}, {and} \bibinfo{person}{Martin Horn}.} \bibinfo{year}{2017}\natexlab{}.
\newblock \showarticletitle{Testing of autonomous vehicles using surrogate models and stochastic optimization}. In \bibinfo{booktitle}{\emph{{ITSC}}}. \bibinfo{publisher}{{IEEE}}, \bibinfo{pages}{1--6}.
\newblock


\bibitem[Biebl et~al\mbox{.}(2021)]%
        {DBLP:conf/hci/BieblKUTRLPB21}
\bibfield{author}{\bibinfo{person}{Bianca Biebl}, \bibinfo{person}{Severin Kacianka}, \bibinfo{person}{Anirudh Unni}, \bibinfo{person}{Alexander Trende}, \bibinfo{person}{Jochem~W. Rieger}, \bibinfo{person}{Andreas L{\"{u}}dtke}, \bibinfo{person}{Alexander Pretschner}, {and} \bibinfo{person}{Klaus Bengler}.} \bibinfo{year}{2021}\natexlab{}.
\newblock \showarticletitle{A Causal Model of Intersection-Related Collisions for Drivers With and Without Visual Field Loss}. In \bibinfo{booktitle}{\emph{{HCI} International 2021 - Late Breaking Papers: {HCI} Applications in Health, Transport, and Industry - 23rd {HCI} International Conference, {HCII} 2021, Virtual Event, July 24-29, 2021 Proceedings}} \emph{(\bibinfo{series}{Lecture Notes in Computer Science}, Vol.~\bibinfo{volume}{13097})}, \bibfield{editor}{\bibinfo{person}{Constantine Stephanidis}, \bibinfo{person}{Vincent~G. Duffy}, \bibinfo{person}{Heidi Kr{\"{o}}mker}, \bibinfo{person}{Fiona~Fui{-}Hoon Nah}, \bibinfo{person}{Keng Siau}, \bibinfo{person}{Gavriel Salvendy}, {and} \bibinfo{person}{June Wei}} (Eds.). \bibinfo{publisher}{Springer}, \bibinfo{pages}{219--234}.
\newblock
\urldef\tempurl%
\url{https://doi.org/10.1007/978-3-030-90966-6\_16}
\showDOI{\tempurl}


\bibitem[Birkemeyer et~al\mbox{.}(2022)]%
        {birkemeyer2022feature}
\bibfield{author}{\bibinfo{person}{Lukas Birkemeyer}, \bibinfo{person}{Tobias Pett}, \bibinfo{person}{Andreas Vogelsang}, \bibinfo{person}{Christoph Seidl}, {and} \bibinfo{person}{Ina Schaefer}.} \bibinfo{year}{2022}\natexlab{}.
\newblock \showarticletitle{Feature-Interaction Sampling for Scenario-based Testing of Advanced Driver Assistance Systems}. In \bibinfo{booktitle}{\emph{Proceedings of the 16th International Working Conference on Variability Modelling of Software-Intensive Systems}}. \bibinfo{pages}{1--10}.
\newblock


\bibitem[{California Department of Motor Vehicles}(2022)]%
        {ca_drivers_handbook}
\bibfield{author}{\bibinfo{person}{{California Department of Motor Vehicles}}.} \bibinfo{year}{2022}\natexlab{}.
\newblock \bibinfo{title}{CA Driver's Handbook}.
\newblock \bibinfo{howpublished}{\url{https://www.dmv.ca.gov/portal/handbook/california-driver-handbook/}}.
\newblock
\newblock
\shownote{Online; accessed December 2022}.


\bibitem[{California Legislative Counsel Bureau}(2022)]%
        {ca_traffic_laws}
\bibfield{author}{\bibinfo{person}{{California Legislative Counsel Bureau}}.} \bibinfo{year}{2022}\natexlab{}.
\newblock \bibinfo{title}{Rules of the Road}.
\newblock \bibinfo{howpublished}{\url{https://leginfo.legislature.ca.gov/faces/codes_displayexpandedbranch.xhtml?tocCode=VEH&division=11.&title=&part=&chapter=&article=&nodetreepath=15}}.
\newblock
\newblock
\shownote{Online; accessed April 2023}.


\bibitem[Correa and Bareinboim(2017)]%
        {correa2017causal}
\bibfield{author}{\bibinfo{person}{Juan Correa} {and} \bibinfo{person}{Elias Bareinboim}.} \bibinfo{year}{2017}\natexlab{}.
\newblock \showarticletitle{Causal effect identification by adjustment under confounding and selection biases}. In \bibinfo{booktitle}{\emph{Proceedings of the AAAI Conference on Artificial Intelligence}}, Vol.~\bibinfo{volume}{31}.
\newblock


\bibitem[Correa et~al\mbox{.}(2018)]%
        {correa2018generalized}
\bibfield{author}{\bibinfo{person}{Juan Correa}, \bibinfo{person}{Jin Tian}, {and} \bibinfo{person}{Elias Bareinboim}.} \bibinfo{year}{2018}\natexlab{}.
\newblock \showarticletitle{Generalized adjustment under confounding and selection biases}. In \bibinfo{booktitle}{\emph{Proceedings of the AAAI Conference on Artificial Intelligence}}, Vol.~\bibinfo{volume}{32}.
\newblock


\bibitem[Das et~al\mbox{.}(2020)]%
        {das2020automated}
\bibfield{author}{\bibinfo{person}{Subasish Das}, \bibinfo{person}{Anandi Dutta}, {and} \bibinfo{person}{Ioannis Tsapakis}.} \bibinfo{year}{2020}\natexlab{}.
\newblock \showarticletitle{Automated vehicle collisions in California: Applying Bayesian latent class model}.
\newblock \bibinfo{journal}{\emph{IATSS research}} \bibinfo{volume}{44}, \bibinfo{number}{4} (\bibinfo{year}{2020}), \bibinfo{pages}{300--308}.
\newblock


\bibitem[Deng et~al\mbox{.}(2022)]%
        {deng2022scenario}
\bibfield{author}{\bibinfo{person}{Yao Deng}, \bibinfo{person}{Xi Zheng}, \bibinfo{person}{Mengshi Zhang}, \bibinfo{person}{Guannan Lou}, {and} \bibinfo{person}{Tianyi Zhang}.} \bibinfo{year}{2022}\natexlab{}.
\newblock \showarticletitle{Scenario-based test reduction and prioritization for multi-module autonomous driving systems}. In \bibinfo{booktitle}{\emph{Proceedings of the 30th ACM Joint European Software Engineering Conference and Symposium on the Foundations of Software Engineering}}. \bibinfo{pages}{82--93}.
\newblock


\bibitem[Dixit et~al\mbox{.}(2016)]%
        {dixit2016autonomous}
\bibfield{author}{\bibinfo{person}{Vinayak~V Dixit}, \bibinfo{person}{Sai Chand}, {and} \bibinfo{person}{Divya~J Nair}.} \bibinfo{year}{2016}\natexlab{}.
\newblock \showarticletitle{Autonomous vehicles: disengagements, accidents and reaction times}.
\newblock \bibinfo{journal}{\emph{PLOS ONE}} \bibinfo{volume}{11}, \bibinfo{number}{12} (\bibinfo{year}{2016}), \bibinfo{pages}{1--14}.
\newblock


\bibitem[Dosovitskiy et~al\mbox{.}(2017)]%
        {dosovitskiy2017carla}
\bibfield{author}{\bibinfo{person}{Alexey Dosovitskiy}, \bibinfo{person}{German Ros}, \bibinfo{person}{Felipe Codevilla}, \bibinfo{person}{Antonio Lopez}, {and} \bibinfo{person}{Vladlen Koltun}.} \bibinfo{year}{2017}\natexlab{}.
\newblock \showarticletitle{CARLA: An open urban driving simulator}. In \bibinfo{booktitle}{\emph{Conference on robot learning}}. PMLR, \bibinfo{pages}{1--16}.
\newblock


\bibitem[Dreossi et~al\mbox{.}(2019a)]%
        {dreossi2019compositional}
\bibfield{author}{\bibinfo{person}{Tommaso Dreossi}, \bibinfo{person}{Alexandre Donz{\'e}}, {and} \bibinfo{person}{Sanjit~A Seshia}.} \bibinfo{year}{2019}\natexlab{a}.
\newblock \showarticletitle{Compositional falsification of cyber-physical systems with machine learning components}.
\newblock \bibinfo{journal}{\emph{Journal of Automated Reasoning}}  \bibinfo{volume}{63} (\bibinfo{year}{2019}), \bibinfo{pages}{1031--1053}.
\newblock


\bibitem[Dreossi et~al\mbox{.}(2019b)]%
        {dreossi2019verifai}
\bibfield{author}{\bibinfo{person}{Tommaso Dreossi}, \bibinfo{person}{Daniel~J Fremont}, \bibinfo{person}{Shromona Ghosh}, \bibinfo{person}{Edward Kim}, \bibinfo{person}{Hadi Ravanbakhsh}, \bibinfo{person}{Marcell Vazquez-Chanlatte}, {and} \bibinfo{person}{Sanjit~A Seshia}.} \bibinfo{year}{2019}\natexlab{b}.
\newblock \showarticletitle{{VerifAI}: A toolkit for the formal design and analysis of artificial intelligence-based systems}. In \bibinfo{booktitle}{\emph{Computer Aided Verification: 31st International Conference, CAV 2019, New York City, NY, USA, July 15-18, 2019, Proceedings, Part I 31}}. Springer, \bibinfo{pages}{432--442}.
\newblock


\bibitem[Fan et~al\mbox{.}(2018)]%
        {fan2018baidu}
\bibfield{author}{\bibinfo{person}{Haoyang Fan}, \bibinfo{person}{Fan Zhu}, \bibinfo{person}{Changchun Liu}, \bibinfo{person}{Liangliang Zhang}, \bibinfo{person}{Li Zhuang}, \bibinfo{person}{Dong Li}, \bibinfo{person}{Weicheng Zhu}, \bibinfo{person}{Jiangtao Hu}, \bibinfo{person}{Hongye Li}, {and} \bibinfo{person}{Qi Kong}.} \bibinfo{year}{2018}\natexlab{}.
\newblock \showarticletitle{{Baidu Apollo EM} Motion Planner}.
\newblock \bibinfo{journal}{\emph{arXiv preprint arXiv:1807.08048}} (\bibinfo{year}{2018}).
\newblock


\bibitem[Favar{\`o} et~al\mbox{.}(2017)]%
        {favaro2017examining}
\bibfield{author}{\bibinfo{person}{Francesca~M Favar{\`o}}, \bibinfo{person}{Nazanin Nader}, \bibinfo{person}{Sky~O Eurich}, \bibinfo{person}{Michelle Tripp}, {and} \bibinfo{person}{Naresh Varadaraju}.} \bibinfo{year}{2017}\natexlab{}.
\newblock \showarticletitle{Examining accident reports involving autonomous vehicles in California}.
\newblock \bibinfo{journal}{\emph{PLoS one}} \bibinfo{volume}{12}, \bibinfo{number}{9} (\bibinfo{year}{2017}), \bibinfo{pages}{e0184952}.
\newblock


\bibitem[Feng et~al\mbox{.}(2021)]%
        {feng2021intelligent}
\bibfield{author}{\bibinfo{person}{Shuo Feng}, \bibinfo{person}{Xintao Yan}, \bibinfo{person}{Haowei Sun}, \bibinfo{person}{Yiheng Feng}, {and} \bibinfo{person}{Henry~X Liu}.} \bibinfo{year}{2021}\natexlab{}.
\newblock \showarticletitle{Intelligent driving intelligence test for autonomous vehicles with naturalistic and adversarial environment}.
\newblock \bibinfo{journal}{\emph{Nature communications}} \bibinfo{volume}{12}, \bibinfo{number}{1} (\bibinfo{year}{2021}), \bibinfo{pages}{748}.
\newblock


\bibitem[Fremont et~al\mbox{.}(2020)]%
        {fremont2020formal}
\bibfield{author}{\bibinfo{person}{Daniel~J Fremont}, \bibinfo{person}{Edward Kim}, \bibinfo{person}{Yash~Vardhan Pant}, \bibinfo{person}{Sanjit~A Seshia}, \bibinfo{person}{Atul Acharya}, \bibinfo{person}{Xantha Bruso}, \bibinfo{person}{Paul Wells}, \bibinfo{person}{Steve Lemke}, \bibinfo{person}{Qiang Lu}, {and} \bibinfo{person}{Shalin Mehta}.} \bibinfo{year}{2020}\natexlab{}.
\newblock \showarticletitle{Formal scenario-based testing of autonomous vehicles: From simulation to the real world}. In \bibinfo{booktitle}{\emph{2020 IEEE 23rd International Conference on Intelligent Transportation Systems (ITSC)}}. IEEE, \bibinfo{pages}{1--8}.
\newblock


\bibitem[Garcia et~al\mbox{.}(2020)]%
        {10.1145/3377811.3380397}
\bibfield{author}{\bibinfo{person}{Joshua Garcia}, \bibinfo{person}{Yang Feng}, \bibinfo{person}{Junjie Shen}, \bibinfo{person}{Sumaya Almanee}, \bibinfo{person}{Yuan Xia}, \bibinfo{person}{Chen}, {and} \bibinfo{person}{Qi Alfred}.} \bibinfo{year}{2020}\natexlab{}.
\newblock \showarticletitle{A Comprehensive Study of Autonomous Vehicle Bugs}. In \bibinfo{booktitle}{\emph{Proceedings of the ACM/IEEE 42nd International Conference on Software Engineering}} (Seoul, South Korea) \emph{(\bibinfo{series}{ICSE '20})}. \bibinfo{publisher}{Association for Computing Machinery}, \bibinfo{address}{New York, NY, USA}, \bibinfo{pages}{385–396}.
\newblock
\showISBNx{9781450371216}
\urldef\tempurl%
\url{https://doi.org/10.1145/3377811.3380397}
\showDOI{\tempurl}


\bibitem[Gonz{\'a}lez et~al\mbox{.}(2015)]%
        {gonzalez2015review}
\bibfield{author}{\bibinfo{person}{David Gonz{\'a}lez}, \bibinfo{person}{Joshu{\'e} P{\'e}rez}, \bibinfo{person}{Vicente Milan{\'e}s}, {and} \bibinfo{person}{Fawzi Nashashibi}.} \bibinfo{year}{2015}\natexlab{}.
\newblock \showarticletitle{A review of motion planning techniques for automated vehicles}.
\newblock \bibinfo{journal}{\emph{IEEE Transactions on intelligent transportation systems}} \bibinfo{volume}{17}, \bibinfo{number}{4} (\bibinfo{year}{2015}), \bibinfo{pages}{1135--1145}.
\newblock


\bibitem[Hauer et~al\mbox{.}(2019)]%
        {hauer2019did}
\bibfield{author}{\bibinfo{person}{Florian Hauer}, \bibinfo{person}{Tabea Schmidt}, \bibinfo{person}{Bernd Holzm{\"{u}}ller}, {and} \bibinfo{person}{Alexander Pretschner}.} \bibinfo{year}{2019}\natexlab{}.
\newblock \showarticletitle{Did We Test All Scenarios for Automated and Autonomous Driving Systems?}. In \bibinfo{booktitle}{\emph{{ITSC}}}. \bibinfo{publisher}{{IEEE}}, \bibinfo{pages}{2950--2955}.
\newblock


\bibitem[Huang et~al\mbox{.}(2017)]%
        {huang2017accelerated}
\bibfield{author}{\bibinfo{person}{Zhiyuan Huang}, \bibinfo{person}{Henry Lam}, {and} \bibinfo{person}{Ding Zhao}.} \bibinfo{year}{2017}\natexlab{}.
\newblock \showarticletitle{An accelerated testing approach for automated vehicles with background traffic described by joint distributions}. In \bibinfo{booktitle}{\emph{2017 IEEE 20th International Conference on Intelligent Transportation Systems (ITSC)}}. IEEE, \bibinfo{pages}{933--938}.
\newblock


\bibitem[Ibrahim et~al\mbox{.}(2019)]%
        {DBLP:conf/nfm/IbrahimKPHK19}
\bibfield{author}{\bibinfo{person}{Amjad Ibrahim}, \bibinfo{person}{Severin Kacianka}, \bibinfo{person}{Alexander Pretschner}, \bibinfo{person}{Charles Hartsell}, {and} \bibinfo{person}{Gabor Karsai}.} \bibinfo{year}{2019}\natexlab{}.
\newblock \showarticletitle{Practical Causal Models for Cyber-Physical Systems}. In \bibinfo{booktitle}{\emph{{NASA} Formal Methods - 11th International Symposium, {NFM} 2019, Houston, TX, USA, May 7-9, 2019, Proceedings}} \emph{(\bibinfo{series}{Lecture Notes in Computer Science}, Vol.~\bibinfo{volume}{11460})}, \bibfield{editor}{\bibinfo{person}{Julia~M. Badger} {and} \bibinfo{person}{Kristin~Yvonne Rozier}} (Eds.). \bibinfo{publisher}{Springer}, \bibinfo{pages}{211--227}.
\newblock
\urldef\tempurl%
\url{https://doi.org/10.1007/978-3-030-20652-9\_14}
\showDOI{\tempurl}


\bibitem[Ibrahim et~al\mbox{.}(2020)]%
        {ibrahim2020actual}
\bibfield{author}{\bibinfo{person}{Amjad Ibrahim}, \bibinfo{person}{Tobias Klesel}, \bibinfo{person}{Ehsan Zibaei}, \bibinfo{person}{Severin Kacianka}, {and} \bibinfo{person}{Alexander Pretschner}.} \bibinfo{year}{2020}\natexlab{}.
\newblock \showarticletitle{Actual causality canvas: a general framework for explanation-based socio-technical constructs}.
\newblock In \bibinfo{booktitle}{\emph{ECAI 2020}}. \bibinfo{publisher}{IOS Press}, \bibinfo{pages}{2978--2985}.
\newblock


\bibitem[Jadidi et~al\mbox{.}(2022)]%
        {JADIDI2022103741}
\bibfield{author}{\bibinfo{person}{Zahra Jadidi}, \bibinfo{person}{Joshua Hagemann}, {and} \bibinfo{person}{Daniel Quevedo}.} \bibinfo{year}{2022}\natexlab{}.
\newblock \showarticletitle{Multi-step attack detection in industrial control systems using causal analysis}.
\newblock \bibinfo{journal}{\emph{Computers in Industry}}  \bibinfo{volume}{142} (\bibinfo{year}{2022}), \bibinfo{pages}{103741}.
\newblock
\showISSN{0166-3615}
\urldef\tempurl%
\url{https://doi.org/10.1016/j.compind.2022.103741}
\showDOI{\tempurl}


\bibitem[Kacianka et~al\mbox{.}(2020)]%
        {DBLP:journals/corr/abs-2005-03294}
\bibfield{author}{\bibinfo{person}{Severin Kacianka}, \bibinfo{person}{Amjad Ibrahim}, {and} \bibinfo{person}{Alexander Pretschner}.} \bibinfo{year}{2020}\natexlab{}.
\newblock \showarticletitle{Expressing Accountability Patterns using Structural Causal Models}.
\newblock \bibinfo{journal}{\emph{CoRR}}  \bibinfo{volume}{abs/2005.03294} (\bibinfo{year}{2020}).
\newblock
\showeprint[arXiv]{2005.03294}
\urldef\tempurl%
\url{https://arxiv.org/abs/2005.03294}
\showURL{%
\tempurl}


\bibitem[Li et~al\mbox{.}(2020)]%
        {li2020av}
\bibfield{author}{\bibinfo{person}{Guanpeng Li}, \bibinfo{person}{Yiran Li}, \bibinfo{person}{Saurabh Jha}, \bibinfo{person}{Timothy Tsai}, \bibinfo{person}{Michael Sullivan}, \bibinfo{person}{Siva Kumar~Sastry Hari}, \bibinfo{person}{Zbigniew Kalbarczyk}, {and} \bibinfo{person}{Ravishankar Iyer}.} \bibinfo{year}{2020}\natexlab{}.
\newblock \showarticletitle{{AV-Fuzzer}: Finding safety violations in autonomous driving systems}. In \bibinfo{booktitle}{\emph{2020 IEEE 31st International Symposium on Software Reliability Engineering (ISSRE)}}. IEEE, \bibinfo{pages}{25--36}.
\newblock


\bibitem[Lou et~al\mbox{.}(2022)]%
        {10.1145/3540250.3549111}
\bibfield{author}{\bibinfo{person}{Guannan Lou}, \bibinfo{person}{Yao Deng}, \bibinfo{person}{Xi Zheng}, \bibinfo{person}{Mengshi Zhang}, {and} \bibinfo{person}{Tianyi Zhang}.} \bibinfo{year}{2022}\natexlab{}.
\newblock \showarticletitle{Testing of Autonomous Driving Systems: Where Are We and Where Should We Go?}. In \bibinfo{booktitle}{\emph{Proceedings of the 30th ACM Joint European Software Engineering Conference and Symposium on the Foundations of Software Engineering}} (Singapore, Singapore) \emph{(\bibinfo{series}{ESEC/FSE 2022})}. \bibinfo{publisher}{Association for Computing Machinery}, \bibinfo{address}{New York, NY, USA}, \bibinfo{pages}{31–43}.
\newblock
\showISBNx{9781450394130}
\urldef\tempurl%
\url{https://doi.org/10.1145/3540250.3549111}
\showDOI{\tempurl}


\bibitem[Luo et~al\mbox{.}(2021)]%
        {luo2021targeting}
\bibfield{author}{\bibinfo{person}{Yixing Luo}, \bibinfo{person}{Xiao-Yi Zhang}, \bibinfo{person}{Paolo Arcaini}, \bibinfo{person}{Zhi Jin}, \bibinfo{person}{Haiyan Zhao}, \bibinfo{person}{Fuyuki Ishikawa}, \bibinfo{person}{Rongxin Wu}, {and} \bibinfo{person}{Tao Xie}.} \bibinfo{year}{2021}\natexlab{}.
\newblock \showarticletitle{Targeting requirements violations of autonomous driving systems by dynamic evolutionary search}. In \bibinfo{booktitle}{\emph{2021 36th IEEE/ACM International Conference on Automated Software Engineering (ASE)}}. IEEE, \bibinfo{pages}{279--291}.
\newblock


\bibitem[Moradi et~al\mbox{.}(2020)]%
        {moradi2020actor}
\bibfield{author}{\bibinfo{person}{Fereidoun Moradi}, \bibinfo{person}{Sara Abbaspour~Asadollah}, \bibinfo{person}{Ali Sedaghatbaf}, \bibinfo{person}{Aida {\v{C}}au{\v{s}}evi{\'c}}, \bibinfo{person}{Marjan Sirjani}, {and} \bibinfo{person}{Carolyn Talcott}.} \bibinfo{year}{2020}\natexlab{}.
\newblock \showarticletitle{An actor-based approach for security analysis of cyber-physical systems}. In \bibinfo{booktitle}{\emph{Formal Methods for Industrial Critical Systems: 25th International Conference, FMICS 2020, Vienna, Austria, September 2--3, 2020, Proceedings 25}}. Springer, \bibinfo{pages}{130--147}.
\newblock


\bibitem[Nigam et~al\mbox{.}(2022)]%
        {nigam_kim_mason_talcott_2022}
\bibfield{author}{\bibinfo{person}{Vivek Nigam}, \bibinfo{person}{Minyoung Kim}, \bibinfo{person}{Ian Mason}, {and} \bibinfo{person}{Carolyn Talcott}.} \bibinfo{year}{2022}\natexlab{}.
\newblock \showarticletitle{Detection and diagnosis of deviations in distributed systems of autonomous agents}.
\newblock \bibinfo{journal}{\emph{Mathematical Structures in Computer Science}} \bibinfo{volume}{32}, \bibinfo{number}{9} (\bibinfo{year}{2022}), \bibinfo{pages}{1254–1282}.
\newblock
\urldef\tempurl%
\url{https://doi.org/10.1017/S0960129522000251}
\showDOI{\tempurl}


\bibitem[Nitsche et~al\mbox{.}(2017)]%
        {nitsche2017pre}
\bibfield{author}{\bibinfo{person}{Philippe Nitsche}, \bibinfo{person}{Pete Thomas}, \bibinfo{person}{Rainer Stuetz}, {and} \bibinfo{person}{Ruth Welsh}.} \bibinfo{year}{2017}\natexlab{}.
\newblock \showarticletitle{Pre-crash scenarios at road junctions: A clustering method for car crash data}.
\newblock \bibinfo{journal}{\emph{Accid. Anal. Prev.}}  \bibinfo{volume}{107} (\bibinfo{year}{2017}), \bibinfo{pages}{137--151}.
\newblock


\bibitem[Paardekooper et~al\mbox{.}(2019)]%
        {paardekooper2019automatic}
\bibfield{author}{\bibinfo{person}{Jan-Pieter Paardekooper}, \bibinfo{person}{S Montfort}, \bibinfo{person}{Jeroen Manders}, \bibinfo{person}{Jorrit Goos}, \bibinfo{person}{E~de Gelder}, \bibinfo{person}{O Camp}, \bibinfo{person}{O Bracquemond}, {and} \bibinfo{person}{Gildas Thiolon}.} \bibinfo{year}{2019}\natexlab{}.
\newblock \showarticletitle{Automatic identification of critical scenarios in a public dataset of 6000 km of public-road driving}. In \bibinfo{booktitle}{\emph{ESV}}.
\newblock


\bibitem[Paden et~al\mbox{.}(2016)]%
        {paden2016survey}
\bibfield{author}{\bibinfo{person}{Brian Paden}, \bibinfo{person}{Michal {\v{C}}{\'a}p}, \bibinfo{person}{Sze~Zheng Yong}, \bibinfo{person}{Dmitry Yershov}, {and} \bibinfo{person}{Emilio Frazzoli}.} \bibinfo{year}{2016}\natexlab{}.
\newblock \showarticletitle{A survey of motion planning and control techniques for self-driving urban vehicles}.
\newblock \bibinfo{journal}{\emph{IEEE Transactions on intelligent vehicles}} \bibinfo{volume}{1}, \bibinfo{number}{1} (\bibinfo{year}{2016}), \bibinfo{pages}{33--55}.
\newblock


\bibitem[Poskitt et~al\mbox{.}(2023)]%
        {Poskitt-et_al23a}
\bibfield{author}{\bibinfo{person}{Christopher~M. Poskitt}, \bibinfo{person}{Yuqi Chen}, \bibinfo{person}{Jun Sun}, {and} \bibinfo{person}{Yu Jiang}.} \bibinfo{year}{2023}\natexlab{}.
\newblock \showarticletitle{Finding Causally Different Tests for an Industrial Control System}. In \bibinfo{booktitle}{\emph{Proc.\ IEEE/ACM International Conference on Software Engineering (ICSE'23)}}. \bibinfo{publisher}{IEEE}, \bibinfo{pages}{2578--2590}.
\newblock


\bibitem[Riccio and Tonella(2020)]%
        {riccio2020model}
\bibfield{author}{\bibinfo{person}{Vincenzo Riccio} {and} \bibinfo{person}{Paolo Tonella}.} \bibinfo{year}{2020}\natexlab{}.
\newblock \showarticletitle{Model-based exploration of the frontier of behaviours for deep learning system testing}. In \bibinfo{booktitle}{\emph{Proceedings of the 28th ACM Joint Meeting on European Software Engineering Conference and Symposium on the Foundations of Software Engineering}}. \bibinfo{pages}{876--888}.
\newblock


\bibitem[Roesener et~al\mbox{.}(2016)]%
        {roesener2016scenario}
\bibfield{author}{\bibinfo{person}{Christian Roesener}, \bibinfo{person}{Felix Fahrenkrog}, \bibinfo{person}{Axel Uhlig}, {and} \bibinfo{person}{Lutz Eckstein}.} \bibinfo{year}{2016}\natexlab{}.
\newblock \showarticletitle{A scenario-based assessment approach for automated driving by using time series classification of human-driving behaviour}. In \bibinfo{booktitle}{\emph{2016 IEEE 19th International Conference on Intelligent Transportation Systems (ITSC)}}. IEEE, \bibinfo{pages}{1360--1365}.
\newblock


\bibitem[Rong et~al\mbox{.}(2020)]%
        {rong2020lgsvl}
\bibfield{author}{\bibinfo{person}{Guodong Rong}, \bibinfo{person}{Byung~Hyun Shin}, \bibinfo{person}{Hadi Tabatabaee}, \bibinfo{person}{Qiang Lu}, \bibinfo{person}{Steve Lemke}, \bibinfo{person}{M{\=a}rti{\c{n}}{\v{s}} Mo{\v{z}}eiko}, \bibinfo{person}{Eric Boise}, \bibinfo{person}{Geehoon Uhm}, \bibinfo{person}{Mark Gerow}, \bibinfo{person}{Shalin Mehta}, {et~al\mbox{.}}} \bibinfo{year}{2020}\natexlab{}.
\newblock \showarticletitle{{LGSVL Simulator}: A high fidelity simulator for autonomous driving}. In \bibinfo{booktitle}{\emph{2020 IEEE 23rd International Conference on Intelligent Transportation Systems (ITSC)}}. IEEE, \bibinfo{pages}{1--6}.
\newblock


\bibitem[Schwarting et~al\mbox{.}(2018)]%
        {schwarting2018planning}
\bibfield{author}{\bibinfo{person}{Wilko Schwarting}, \bibinfo{person}{Javier Alonso-Mora}, {and} \bibinfo{person}{Daniela Rus}.} \bibinfo{year}{2018}\natexlab{}.
\newblock \showarticletitle{Planning and decision-making for autonomous vehicles}.
\newblock \bibinfo{journal}{\emph{Annual Review of Control, Robotics, and Autonomous Systems}}  \bibinfo{volume}{1} (\bibinfo{year}{2018}), \bibinfo{pages}{187--210}.
\newblock


\bibitem[Shalev{-}Shwartz et~al\mbox{.}(2017)]%
        {DBLP:journals/corr/abs-1708-06374}
\bibfield{author}{\bibinfo{person}{Shai Shalev{-}Shwartz}, \bibinfo{person}{Shaked Shammah}, {and} \bibinfo{person}{Amnon Shashua}.} \bibinfo{year}{2017}\natexlab{}.
\newblock \showarticletitle{On a Formal Model of Safe and Scalable Self-driving Cars}.
\newblock \bibinfo{journal}{\emph{CoRR}}  \bibinfo{volume}{abs/1708.06374} (\bibinfo{year}{2017}).
\newblock
\showeprint[arXiv]{1708.06374}
\urldef\tempurl%
\url{http://arxiv.org/abs/1708.06374}
\showURL{%
\tempurl}


\bibitem[Sun et~al\mbox{.}(2022)]%
        {sun2022lawbreaker}
\bibfield{author}{\bibinfo{person}{Yang Sun}, \bibinfo{person}{Christopher~M Poskitt}, \bibinfo{person}{Jun Sun}, \bibinfo{person}{Yuqi Chen}, {and} \bibinfo{person}{Zijiang Yang}.} \bibinfo{year}{2022}\natexlab{}.
\newblock \showarticletitle{LawBreaker: An Approach for Specifying Traffic Laws and Fuzzing Autonomous Vehicles}. In \bibinfo{booktitle}{\emph{37th IEEE/ACM International Conference on Automated Software Engineering}}. \bibinfo{pages}{1--12}.
\newblock


\bibitem[Tokdar and Kass(2010)]%
        {tokdar2010importance}
\bibfield{author}{\bibinfo{person}{Surya~T Tokdar} {and} \bibinfo{person}{Robert~E Kass}.} \bibinfo{year}{2010}\natexlab{}.
\newblock \showarticletitle{Importance sampling: a review}.
\newblock \bibinfo{journal}{\emph{Wiley Interdisciplinary Reviews: Computational Statistics}} \bibinfo{volume}{2}, \bibinfo{number}{1} (\bibinfo{year}{2010}), \bibinfo{pages}{54--60}.
\newblock


\bibitem[Wan et~al\mbox{.}(2022)]%
        {ndss:2022:ziwen:planfuzz}
\bibfield{author}{\bibinfo{person}{Ziwen Wan}, \bibinfo{person}{Junjie Shen}, \bibinfo{person}{Jalen Chuang}, \bibinfo{person}{Xin Xia}, \bibinfo{person}{Joshua Garcia}, \bibinfo{person}{Jiaqi Ma}, {and} \bibinfo{person}{Qi~Alfred Chen}.} \bibinfo{year}{2022}\natexlab{}.
\newblock \showarticletitle{{Too Afraid to Drive: Systematic Discovery of Semantic DoS Vulnerability in Autonomous Driving Planning under Physical-World Attacks}}. In \bibinfo{booktitle}{\emph{Network and Distributed System Security (NDSS) Symposium, 2022}}.
\newblock


\bibitem[Wang and Li(2019)]%
        {wang2019exploring}
\bibfield{author}{\bibinfo{person}{Song Wang} {and} \bibinfo{person}{Zhixia Li}.} \bibinfo{year}{2019}\natexlab{}.
\newblock \showarticletitle{Exploring the mechanism of crashes with automated vehicles using statistical modeling approaches}.
\newblock \bibinfo{journal}{\emph{PloS one}} \bibinfo{volume}{14}, \bibinfo{number}{3} (\bibinfo{year}{2019}), \bibinfo{pages}{e0214550}.
\newblock


\bibitem[Wang et~al\mbox{.}(2021)]%
        {wang2021combining}
\bibfield{author}{\bibinfo{person}{Xinpeng Wang}, \bibinfo{person}{Huei Peng}, {and} \bibinfo{person}{Ding Zhao}.} \bibinfo{year}{2021}\natexlab{}.
\newblock \showarticletitle{Combining reachability analysis and importance sampling for accelerated evaluation of highway automated vehicles at pedestrian crossing}.
\newblock \bibinfo{journal}{\emph{ASME Letters in Dynamic Systems and Control}} \bibinfo{volume}{1}, \bibinfo{number}{1} (\bibinfo{year}{2021}).
\newblock


\bibitem[Werling et~al\mbox{.}(2010)]%
        {werling2010optimal}
\bibfield{author}{\bibinfo{person}{Moritz Werling}, \bibinfo{person}{Julius Ziegler}, \bibinfo{person}{S{\"o}ren Kammel}, {and} \bibinfo{person}{Sebastian Thrun}.} \bibinfo{year}{2010}\natexlab{}.
\newblock \showarticletitle{Optimal trajectory generation for dynamic street scenarios in a frenet frame}. In \bibinfo{booktitle}{\emph{2010 IEEE International Conference on Robotics and Automation}}. IEEE, \bibinfo{pages}{987--993}.
\newblock


\bibitem[Zhang et~al\mbox{.}(2014)]%
        {zhang2014roadview}
\bibfield{author}{\bibinfo{person}{Chi Zhang}, \bibinfo{person}{Yuehu Liu}, \bibinfo{person}{Danchen Zhao}, {and} \bibinfo{person}{Yuanqi Su}.} \bibinfo{year}{2014}\natexlab{}.
\newblock \showarticletitle{{RoadView}: A traffic scene simulator for autonomous vehicle simulation testing}. In \bibinfo{booktitle}{\emph{17th International IEEE Conference on Intelligent Transportation Systems (ITSC)}}. IEEE, \bibinfo{pages}{1160--1165}.
\newblock


\bibitem[Zhang and Bareinboim(2017)]%
        {zhang2017transfer}
\bibfield{author}{\bibinfo{person}{Junzhe Zhang} {and} \bibinfo{person}{Elias Bareinboim}.} \bibinfo{year}{2017}\natexlab{}.
\newblock \showarticletitle{Transfer learning in multi-armed bandit: a causal approach}. In \bibinfo{booktitle}{\emph{Proceedings of the 16th Conference on Autonomous Agents and MultiAgent Systems}}. \bibinfo{pages}{1778--1780}.
\newblock


\bibitem[Zhang et~al\mbox{.}(2020)]%
        {zhang2020tracing}
\bibfield{author}{\bibinfo{person}{Renbin Zhang}, \bibinfo{person}{Zongze Cao}, {and} \bibinfo{person}{Kewei Wu}.} \bibinfo{year}{2020}\natexlab{}.
\newblock \showarticletitle{Tracing and detection of ICS anomalies based on causality mutations}. In \bibinfo{booktitle}{\emph{2020 IEEE 5th Information Technology and Mechatronics Engineering Conference (ITOEC)}}. IEEE, \bibinfo{pages}{511--517}.
\newblock


\bibitem[Zhao et~al\mbox{.}(2017)]%
        {zhao2017accelerated}
\bibfield{author}{\bibinfo{person}{Ding Zhao}, \bibinfo{person}{Xianan Huang}, \bibinfo{person}{Huei Peng}, \bibinfo{person}{Henry Lam}, {and} \bibinfo{person}{David~J LeBlanc}.} \bibinfo{year}{2017}\natexlab{}.
\newblock \showarticletitle{Accelerated evaluation of automated vehicles in car-following maneuvers}.
\newblock \bibinfo{journal}{\emph{IEEE Transactions on Intelligent Transportation Systems}} \bibinfo{volume}{19}, \bibinfo{number}{3} (\bibinfo{year}{2017}), \bibinfo{pages}{733--744}.
\newblock


\bibitem[Zhao et~al\mbox{.}(2016)]%
        {zhao2016accelerated}
\bibfield{author}{\bibinfo{person}{Ding Zhao}, \bibinfo{person}{Henry Lam}, \bibinfo{person}{Huei Peng}, \bibinfo{person}{Shan Bao}, \bibinfo{person}{David~J LeBlanc}, \bibinfo{person}{Kazutoshi Nobukawa}, {and} \bibinfo{person}{Christopher~S Pan}.} \bibinfo{year}{2016}\natexlab{}.
\newblock \showarticletitle{Accelerated evaluation of automated vehicles safety in lane-change scenarios based on importance sampling techniques}.
\newblock \bibinfo{journal}{\emph{IEEE Transactions on Intelligent Transportation Systems}} \bibinfo{volume}{18}, \bibinfo{number}{3} (\bibinfo{year}{2016}), \bibinfo{pages}{595--607}.
\newblock


\bibitem[Zhou et~al\mbox{.}(2023)]%
        {zhou2023specification}
\bibfield{author}{\bibinfo{person}{Yuan Zhou}, \bibinfo{person}{Yang Sun}, \bibinfo{person}{Yun Tang}, \bibinfo{person}{Yuqi Chen}, \bibinfo{person}{Jun Sun}, \bibinfo{person}{Christopher~M Poskitt}, \bibinfo{person}{Yang Liu}, {and} \bibinfo{person}{Zijiang Yang}.} \bibinfo{year}{2023}\natexlab{}.
\newblock \showarticletitle{Specification-based Autonomous Driving System Testing}.
\newblock \bibinfo{journal}{\emph{IEEE Transactions on Software Engineering}} \bibinfo{volume}{49}, \bibinfo{number}{6} (\bibinfo{year}{2023}), \bibinfo{pages}{3391--3410}.
\newblock


\bibitem[Zohdinasab et~al\mbox{.}(2021)]%
        {zohdinasab2021deephyperion}
\bibfield{author}{\bibinfo{person}{Tahereh Zohdinasab}, \bibinfo{person}{Vincenzo Riccio}, \bibinfo{person}{Alessio Gambi}, {and} \bibinfo{person}{Paolo Tonella}.} \bibinfo{year}{2021}\natexlab{}.
\newblock \showarticletitle{Deephyperion: exploring the feature space of deep learning-based systems through illumination search}. In \bibinfo{booktitle}{\emph{Proceedings of the 30th ACM SIGSOFT International Symposium on Software Testing and Analysis}}. \bibinfo{pages}{79--90}.
\newblock


\end{thebibliography}

\end{document}